%% file: main.tex
\input{preamble}
\input{authors}%
\begin{document}


\maketitle
\input{frontmatter}%



\section{Introduction}
Many fundamental graph optimization problems remain NP-hard even on highly structured inputs. Three complementary responses to this difficulty are to seek fixed-parameter algorithms, to settle for approximate solutions, and to exploit restrictions on the input graph. The interaction between these approaches has been particularly fruitful on planar graphs and, more generally, on graphs excluding a fixed minor. Bidimensionality provides a unified explanation for much of this tractability: for broad families of problems, suitable structural and algorithmic hypotheses yield subexponential parameterized algorithms exhibiting the square-root phenomenon (the running time dependence on the parameter $k$ is about $2^{O(\sqrt{k})}$), efficient polynomial-time approximation schemes (EPTASes), and linear kernels~\cite{DemaineFominHajiaghayiThilikos2005,DemaineHajiaghayi2008,FominLokshtanovRamanSaurabh2011,FominLokshtanovSaurabhThilikos2020}. A key structural ingredient in the planar case is the linear relation between treewidth and the size of a largest grid minor. 

There are two natural directions in which to look. One is to consider more general sparse graph classes; another is to consider geometric intersection graphs, including unit disk graphs, disk graphs, rectangle intersection graphs, and string graphs. The first direction already encounters a basic obstruction: \textsc{Vertex Cover} is APX-complete on cubic graphs~\cite{AlimontiKann2000}. Since bounded-degree classes have bounded expansion and exclude suitable fixed topological minors, these restrictions alone cannot support a general approximation-scheme theorem unless $\mathrm{P}=\mathrm{NP}$. Geometric classes offer a different source of structure. They may contain arbitrarily large cliques, yet still admit useful separator and decomposition theorems. De Berg et al.~\cite{deBergBodlaenderKisfaludiBakMarxVanDerZanden2018} developed a general framework for subexponential algorithms on intersection graphs of similarly sized fat objects, based on geometric separators and treewidth techniques. Fomin, Lokshtanov, and Saurabh~\cite{FominLokshtanovSaurabh2012Geometric} extended bidimensionality techniques to obtain EPTASes and subexponential parameterized algorithms for several deletion problems on unit disk and map graphs. Further subexponential parameterized algorithms on unit disk and disk graphs illustrate the potential of this approach~\cite{AnOh2021,LokshtanovPanolanSaurabhXueZehavi2022,AnChoOh2023}. Lokshtanov et al.~\cite{LokshtanovPanolanSaurabhXueZehavi2023ApproxDisk} subsequently developed a representation-free EPTAS framework for deletion problems on disk graphs, including \textsc{Vertex Cover} and \textsc{Feedback Vertex Set}.

\textsc{Feedback Vertex Set} has been a central test case for these extensions. On unit disk graphs, Fomin et al.~\cite{FominLokshtanovPanolanSaurabhZehavi2017} obtained a $2^{O(\sqrt{k}\log k)}n^{O(1)}$ algorithm, and An and Oh~\cite{AnOh2021} improved this to $2^{O(\sqrt{k})}(n+m)$ without requiring a representation. Baste and Thilikos~\cite{BasteThilikos2022} extended bidimensionality to further geometric classes, including bounded-degree string graphs. Subsequent work developed algorithms for disks~\cite{AnChoOh2023}, squares and contact segments~\cite{BertheBougeretGoncalvesRaymond2024LocalRadius}, and pseudo-disks~\cite{BertheBougeretGoncalvesRaymond2024PseudoDiskWG,BertheBougeretGoncalvesRaymond2024PseudoDiskFull}.

Recent work has developed broader frameworks beyond these classes. Berthe et al.~\cite{BertheBougeretGoncalvesRaymond2025} give sufficient conditions for subexponential parameterized algorithms for \textsc{Feedback Vertex Set}, covering, in particular, segment graphs and $s$-string graphs for fixed $s$. Lokshtanov et al.~\cite{LokshtanovPanolanSaurabhXueZehavi2025Hitting} give a framework for hitting a fixed finite family of forbidden subgraphs on classes with suitable separator bounds, including classes of polynomial expansion and several geometric classes. A fixed finite family of forbidden subgraphs does not capture \textsc{Feedback Vertex Set}, since cycles can have unbounded length. Our bounded-treewidth target formulation accommodates this problem together with more general MSO constraints.

\begin{figure}[tbp]
\centering
\begin{tikzpicture}[
  font=\small,
  graphclass/.style={draw=black!65, rounded corners=3pt, fill=black!3,
    align=center, text width=2.8cm, minimum height=0.8cm, inner sep=5pt},
  inclusion/.style={-{Stealth[length=2mm]}, thick, draw=black!75},
  covered/.style={inclusion, dashed},
  annotation/.style={font=\footnotesize, fill=white, inner sep=2pt}
]
\node[graphclass] (unit) at (0,0) {Unit disk graphs};
\node[graphclass] (planar) at (-5,-1.5) {Planar graphs};
\node[graphclass] (disk) at (0,-1.5) {Disk graphs};
\node[graphclass] (rectangle) at (5,-1.5) {Rectangle intersection\\graphs};
\node[graphclass] (minor) at (-5,-3.3) {$H$-minor-free\\graphs};
\node[graphclass] (string) at (2.5,-3.3) {String graphs};
\node[draw=blue!55!black, line width=1pt, rounded corners=4pt,
  fill=blue!5, align=center, text width=13.4cm, inner sep=8pt]
  (framework) at (0,-5.4) {
    \textbf{Our framework: $H$-induced-minor-free graphs, for every fixed $H$}\\[3pt]
    Bounded-treewidth MSO vertex deletion:\\[2pt]
    subexponential FPT algorithms; QPTASes for hereditary target properties
  };
\draw[inclusion] (unit) -- (disk);
\draw[inclusion] (planar) -- node[annotation,left] {$H$ nonplanar} (minor);
\draw[inclusion] (planar.east) -- (string.west);
\draw[inclusion] (disk.south) -- (string.north west);
\draw[inclusion] (rectangle.south) -- (string.north east);
\draw[covered] (minor.south) -- (framework.north -| minor.south);
\draw[covered] (string.south) -- (framework.north -| string.south);
\end{tikzpicture}
\caption{Graph classes covered by our framework. Solid arrows indicate containment; only selected containments are shown. Dashed arrows indicate applications of the framework, with the excluded graph chosen for the class: the same $H$ for $H$-minor-free graphs, and the once-subdivided $K_5$ for string graphs~\cite{korhonen2024induced}. The highlighted box summarizes Theorems~\ref{thm:subexp} and~\ref{thm:qptas}; no geometric representation is required.}
\label{fig:graph-class-scope}
\end{figure}
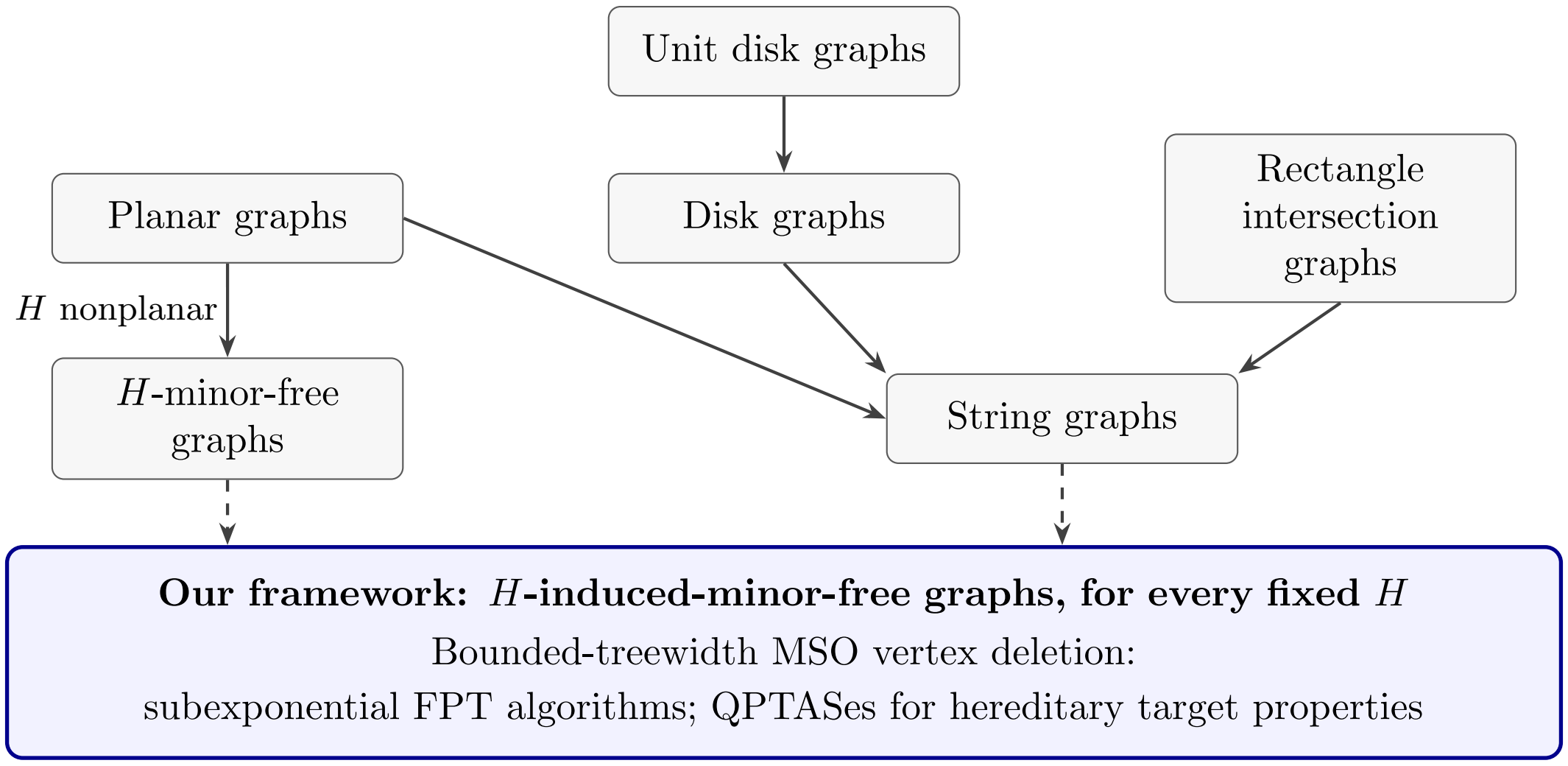

\paragraph{From string graphs to excluded induced minors.}
A string graph is the intersection graph of a family of curves in the plane. This class includes disk and rectangle intersection graphs, as well as all planar graphs. A useful common setting is provided by graphs excluding a fixed graph as an \emph{induced minor}, where an induced minor is obtained by vertex deletions and edge contractions. For a fixed graph $H$, let $\mathcal{G}_H$ denote the class of $H$-induced-minor-free graphs. Every string graph excludes the graph obtained by subdividing each edge of $K_5$ once as an induced minor~\cite{korhonen2024induced}. Thus, an algorithmic theorem for every fixed $H$ applies, in particular, to all string graphs. Figure~\ref{fig:graph-class-scope} summarizes the graph classes covered by this framework.

The relevant structural ingredient is an edge-based separator theorem: graphs in $\mathcal{G}_H$ with $m$ edges have balanced separators of size $O_H(\sqrt{m})$. Lee~\cite{Lee2017} and Korhonen and Lokshtanov~\cite{korhonen2024induced} originally claimed these bounds, but their proofs relied on a flawed padded-decomposition argument. Davies, Rz{\fontencoding{T1}\selectfont\k{a}}\.{z}ewski, Spalding-Jamieson, and Walczak~\cite{DaviesRzazewskiSpaldingJamiesonWalczak2026} give new proofs and polynomial-time separator algorithms. Both of our main results, Theorems~\ref{thm:subexp} and~\ref{thm:qptas}, rely on this repaired separator theorem. It also restores the separator input underlying the $2^{\widetilde{O}(n^{2/3})}$ algorithms of Korhonen and Lokshtanov. Unlike the separators of sparse minor-free graphs, these bounds are expressed in terms of the number of edges. Controlling dense parts of the input is therefore essential to obtaining algorithms whose complexity is governed by the solution size.

\paragraph{The deletion framework.}
We study vertex-deletion problems to classes of graphs with bounded treewidth and satisfying a fixed logical property. More precisely, fix an integer $t$ and a monadic second-order (MSO) sentence $\phi$. The objective is to delete as few vertices as possible so that the remaining graph has treewidth at most $t$ and satisfies $\phi$. This formulation includes \textsc{Vertex Cover}, by requiring the remaining graph to be edgeless; \textsc{Feedback Vertex Set}, by taking $t=1$ and imposing no further condition; and deletion to treewidth at most $t$. For approximation, we require the property defined by $\phi$ to be hereditary, meaning that it is preserved under taking induced subgraphs. The problem is stated formally below (where $t \in \mathbb{N}$ and $\phi$ is an MSO formula).

\begin{tcolorbox}[colback=gray!5!white,colframe=gray!75!black]
        \textsc{$(t,\phi)$-Deletion} \hfill \textbf{Parameter:} $k = |S|$
        \vspace{0.2cm} \\
        \textbf{Input:} A graph $G$ with $n$ vertices
        \vspace{0.1cm} \\
        \textbf{Output:} A minimum subset $S \subseteq V(G)$ satisfying $\mathbf{tw}(G-S) \leq t$ and $G-S$ satisfies $\phi$
\end{tcolorbox}

A very closely related class of problems is \textsc{Planar $\mathcal{F}$-Deletion}: delete vertices such that the resulting graph excludes every graph in a fixed finite family $\mathcal{F}$ as a minor. The family $\mathcal{F}$ is required to contain at least one planar graph. 
It follows from the excluded grid theorem of Robertson and Seymour~\cite{RobertsonSeymour1986Grid} that every graph excluding a fixed planar graph as a minor has treewidth bounded by a constant; see also Cames van Batenburg et al.~\cite{CamesVanBatenburgClausJoretPetitRaymondRobinson2026Grid} for a recent improvement of the grid theorem for planar graphs. Further, containing a fixed graph as a minor is MSO definable. Thus, for every finite family $\mathcal{F}$ containing a planar graph, there exists an integer $t$ and MSO formula $\phi$ such that \textsc{Planar $\mathcal{F}$-Deletion} is a \textsc{$(t,\phi)$-Deletion} problem.
Fomin et al.~\cite{FominLokshtanovMisraSaurabh2012} established constant-factor approximation algorithms and polynomial kernels for \textsc{Planar $\mathcal{F}$-Deletion}, as well as single-exponential parameterized algorithms when every graph in $\mathcal{F}$ is connected. Kim et al.~\cite{KimLangerPaulReidlRossmanithSauSikdar2015} removed this connectivity assumption for the single-exponential algorithm. Subsequent work of Gupta et al.~\cite{DBLP:conf/soda/GuptaLLM019} gave deterministic approximation algorithms with improved guarantees. More recently, W{\l}odarczyk~\cite{Wlodarczyk2025LosingWeights} obtained a randomized polynomial-time constant-factor approximation algorithm for the vertex-weighted version.  

Our focus is the additional algorithmic power obtained by restricting the input to $\mathcal{G}_H$.

\paragraph{Our results.}
Our first main result is a subexponential parameterized algorithm for every \textsc{$(t,\phi)$-Deletion} problem. We use $k$ for the size of an optimum deletion set; the notation $\widetilde{O}$ suppresses polylogarithmic factors. 

\begin{restatable}{theorem}{subexp} \label{thm:subexp}
    Let $H$ be a fixed graph, $t \in \mathbb{N}$ be a fixed number, and $\phi$ be a fixed MSO formula.
    Then \textsc{$(t,\phi)$-Deletion} on $\mathcal{G}_H$ admits an algorithm that runs in $2^{\widetilde{O}(k^{2/3})} n^{O(1)}$ time.
\end{restatable}

The technical framework behind Theorem~\ref{thm:subexp} also yields quasipolynomial-time approximation schemes for almost every \textsc{$(t,\phi)$-Deletion} problem, which is our second main result. The only additional property that we require is that the target graph class is closed under vertex deletion. 

\begin{restatable}{theorem}{qptas} \label{thm:qptas}
    Let $H$ be a fixed graph, $t \in \mathbb{N}$ be a fixed number, and $\phi$ be a fixed hereditary MSO formula.
    Then \textsc{$(t,\phi)$-Deletion} on $\mathcal{G}_H$ admits a quasipolynomial-time approximation scheme.
\end{restatable}

Theorem~\ref{thm:subexp} measures the exponential part of the running time in the number $k$ of deleted vertices. The size $k$ of the solution is always at most the number $n$ of vertices in the input graph, and it can be much smaller. Korhonen and Lokshtanov~\cite{korhonen2024induced}, generalizing earlier work by Bonnet and Rz{\fontencoding{T1}\selectfont\k{a}}\.{z}ewski~\cite{BonnetRzazewski2019}, obtained $2^{\widetilde{O}(n^{2/3})}$-time algorithms for a broad family of induced-subgraph problems that includes every \textsc{$(t,\phi)$-Deletion} problem.
%
The running time bound of Theorem~\ref{thm:subexp} improves over the running time bounds for {\em more restricted} classes of problems on {\em more restricted} classes of graphs (such as Disk Graphs)
~\cite{AnChoOh2023,BertheBougeretGoncalvesRaymond2024PseudoDiskFull,BertheBougeretGoncalvesRaymond2024LocalRadius,BertheBougeretGoncalvesRaymond2025}, see Table~\ref{tab:fvs-comparison}.
It answers in the affirmative an open problem in Section~6 of the full version of~\cite{BertheBougeretGoncalvesRaymond2025}: obtaining a subexponential algorithm for \textsc{Feedback Vertex Set} parameterized by $k$ on classes excluding a fixed induced minor, and in particular on string graphs.

The running time dependence on $k$ of Theorem~\ref{thm:subexp} essentially matches the running time dependence on $n$ of Korhonen and Lokshtanov~\cite{korhonen2024induced}. Thus any further improvement to the running time of Theorem~\ref{thm:subexp} must be preceeded by faster algorithms for these problems in terms of the number of vertices. 

%
%
%
%

Theorem~\ref{thm:qptas} states that, for every fixed $\varepsilon>0$ and hereditary $\phi$, a deletion set of size at most $(1+\varepsilon)k$ can be found in quasipolynomial time. Both results apply to the graph itself and do not require a geometric representation.
For \textsc{Vertex Cover} and \textsc{Feedback Vertex Set} on disk graphs, representation-free EPTASes were already obtained by Lokshtanov et al.~\cite{LokshtanovPanolanSaurabhXueZehavi2023ApproxDisk}. Our approximation contribution for these problems is therefore the extension to broader classes.


\begin{table}[tbp]
\centering
\small
\renewcommand{\arraystretch}{1.18}
\setlength{\tabcolsep}{5pt}
\begin{tabular}{l c c c}
\hline
\textbf{Graph class} & \textbf{Representation} & \textbf{Earlier $\alpha$} & \textbf{Our $\alpha$} \\
 & \textbf{needed before?} & & \\
\hline
Disk graphs & No & $9/10$~\cite{AnChoOh2023} & $2/3$ \\
Disk graphs & Yes & $7/8$~\cite{AnChoOh2023} & $2/3$ \\
Pseudo-disk graphs & Yes & $6/7$~\cite{BertheBougeretGoncalvesRaymond2024PseudoDiskFull} & $2/3$ \\
Pseudo-disk graphs & No & $10/11$~\cite{BertheBougeretGoncalvesRaymond2024PseudoDiskFull} & $2/3$ \\
Axis-parallel square graphs & No & $9/10$~\cite{BertheBougeretGoncalvesRaymond2024LocalRadius} & $2/3$ \\
$s$-string graphs, fixed $s$ & No & $52/53$~\cite{BertheBougeretGoncalvesRaymond2025} & $2/3$ \\
All string graphs & --- & Open~\cite{BertheBougeretGoncalvesRaymond2025} & $2/3$ \\
\hline
\end{tabular}
\caption{Selected comparisons for \textsc{Feedback Vertex Set}. An exponent $\alpha$ denotes a bound $2^{\widetilde{O}(k^\alpha)}n^{O(1)}$; all bounds from this paper require only the input graph. The pseudo-disk bounds $6/7$ and $10/11$ are from arXiv:2410.23878v1~\cite{BertheBougeretGoncalvesRaymond2024PseudoDiskFull}; the published WG version~\cite{BertheBougeretGoncalvesRaymond2024PseudoDiskWG} states the earlier represented bound $9/10$.
The $52/53$ bound is explicit in Section~5.2 of arXiv:2504.17708v2~\cite{BertheBougeretGoncalvesRaymond2025}. The $6/7$ pseudo-disk bound also applies to represented disk graphs. \emph{Open} refers to a subexponential algorithm parameterized by $k$, not by $n$. 
}
\label{tab:fvs-comparison}
\end{table}

\subsection{Related Work on Approximation Algorithms}
We next place our approximation result in the context of earlier work on geometric intersection graphs and excluded-induced-minor classes. We first discuss \textsc{Vertex Cover}, and then approximation schemes for finding large independent sets and other induced subgraphs.

\par\medskip\noindent\textbf{Approximation of Vertex Cover on geometric graphs.}\enspace
The approximation literature also goes well beyond the factor-$2$ guarantee available on general graphs. Erlebach, Jansen, and Seidel~\cite{ErlebachJansenSeidel2005} gave PTASes for weighted \textsc{Vertex Cover} on intersection graphs of fat objects. For axis-parallel rectangles, Bar-Yehuda, Hermelin, and Rawitz~\cite{BarYehudaHermelinRawitz2011} obtained a $(3/2+\varepsilon)$-approximation, also in the weighted case. Bar-Yehuda et al.~\cite{BarYehudaHermelinRawitz2011} also gave an EPTAS for unweighted \textsc{Vertex Cover} on pseudo-disk graphs. The unweighted independent-set PTAS of Chan and Har-Peled~\cite{ChanHarPeled2012}, combined with the transfer discussed by Har-Peled~\cite{HarPeled2023IndependenceCover}, yields a PTAS for \textsc{Vertex Cover} on pseudo-disk graphs. These results concern subclasses of string graphs.

For general string graphs, Lokshtanov et al.~\cite{LokshtanovPanolanSaurabhXueZehavi2024VC} gave a $1.9999$-approximation, including a weighted version. Its original proof relied on a flawed decomposition result of Lee, as explained by Bonnet and Rz{\fontencoding{T1}\selectfont\k{a}}{\.{z}}ewski~\cite{BonnetRzazewski2025VC}. Davies et al.~\cite[Theorem~1.4 and Section~8]{DaviesRzazewskiSpaldingJamiesonWalczak2026} repair the required decomposition theorems for string graphs, without requiring a representation, thereby restoring the weighted $1.9999$-approximation. The claimed extension of that argument to all fixed excluded-induced-minor classes remains unresolved. Bonnet and Rz{\fontencoding{T1}\selectfont\k{a}}{\.{z}}ewski~\cite{BonnetRzazewski2025VC} obtained an $11/6$-approximation for unweighted \textsc{Vertex Cover} in time polynomial in the size of a supplied string representation, as well as a representation-free polynomial-time $11/6$ gap-decision algorithm. Thus the representation model and the distinction between producing a cover and solving a gap-decision problem must be kept explicit. These polynomial-time constant-factor results complement the quasipolynomial-time schemes discussed next.

\par\medskip\noindent\textbf{QPTASes for geometric packing and retained subgraphs.}\enspace
The QPTAS of Adamaszek and Wiese~\cite{AdamaszekWiese2013} for maximum-weight independent set of axis-parallel rectangles initiated a line of geometric approximation schemes. Adamaszek, Har-Peled, and Wiese~\cite{AdamaszekHarPeledWiese2019} extended this approach to polygons and to finding large sparse induced subgraphs of their intersection graphs. Pilipczuk, van Leeuwen, and Wiese~\cite{PilipczukVanLeeuwenWiese2020} developed a combinatorial formulation through packing connected subgraphs of a planar host. These geometric algorithms use the corresponding representations. In a different direction, Dvo\v{r}{\'a}k~\cite{Dvorak2018Separators} and Dvo\v{r}{\'a}k and Lahiri~\cite{DvorakLahiri2021} obtained QPTASes for weighted independent set and related problems under strongly sublinear vertex-separator assumptions. Dvo\v{r}{\'a}k~\cite{Dvorak2022Approximation} subsequently established approximation metatheorems for monotone first-order maximization problems. For induced-minor restrictions, Bonnet et al.~\cite{BonnetCzyzewskaMasarikPilipczukRzazewski2026} obtained QPTASes on $sC_\ell$-induced-minor-free graphs, for fixed $s,\ell$, both for weighted independent set and for maximizing the order of a bounded-treewidth induced subgraph satisfying a fixed CMSO$_2$ property closed under induced subgraphs and disjoint unions; the latter includes \textsc{Maximum Induced Forest}.

The recent theorem of Bonnet and Chang~\cite[Theorem~2]{BonnetChang2026} gives a deterministic QPTAS for \textsc{Maximum Weight Independent Set} on every hereditary class with balanced separators of size $O(1+m^b)$, for a fixed $b<1$, where $m$ is the number of edges. In particular, it applies to $\mathcal{G}_H$ for every fixed $H$. Even though they do not state this explicitly, their approximation schemes very likely extend to the {\em maximization version} of \textsc{$(t,\phi)$-Deletion}.

In general, an approximation scheme for the maximization version of a problem does not imply an approximation scheme for the minimization version of the same problem, or vice versa. Thus the results in Theorem~\ref{thm:qptas} and the algorithm of Bonnet and Chang are incomparable. A notable exception is for {\sc Vertex Cover}, for which a QPTAS {\em does} follow from the work of  Bonnet and Chang~\cite{BonnetChang2026} by the weighted Nemhauser--Trotter reduction, in the form discussed by Har-Peled~\cite{HarPeled2023IndependenceCover}.

\paragraph{Organization.}
The rest of the paper is organized as follows.
Section~\ref{sec:overview} gives an overview of the main ideas behind our proofs and Section~3 introduces the notation and preliminary tools.
In Section~4, we prove the sublinear-treewidth branching theorem, which is the technical core of both the subexponential FPT algorithm and the QPTAS.
Section~5 presents a dynamic programming algorithm for solving \textsc{$(t,\phi)$-Deletion} parameterized by treewidth.
Sections~\ref{sec:subexp} and~\ref{sec:QPTAS} combine these tools to prove the subexponential FPT algorithm and the QPTAS, respectively.

\section{Technical overview}
\label{sec:overview}

For simplicity, we overview our algorithms for the \textsc{Feedback Vertex Set} (FVS) problem, which is the special case of \textsc{$(t,\phi)$-Deletion} when $t=1$ and $\phi$ is the dummy formula.
Considering this special case is already sufficient to get most of the insights in our results.

Our subexponential FPT algorithm and QPTAS share the same technical core, which we call the \textit{sublinear-treewidth branching theorem} (Theorem~\ref{thm:branching}).
Essentially, this theorem allows us to ``guess'' in subexponential FPT time a subset $X$ of the unknown solution $S \subseteq V(G)$ such that the treewidth of $G-X$ is sublinear in $k$.
Below is the special case of the theorm for FVS.


\medskip

\noindent
\textbf{Sublinear-treewidth branching (STB) theorem.}
{\it Given an $H$-induced-minor-free graph $G$ with $n$ vertices and two parameters $k,d \in \mathbb{N}$, one can compute in $2^{\widetilde{O}(k/d)} n^{O(1)}$ time a collection $\mathcal{X} = \{X_1,\dots,X_r\}$ of subsets of $V(G)$, where $r = 2^{\widetilde{O}(k/d)}$, such that
\begin{enumerate}[(i)]
    \item $\mathbf{tw}(G-X_i) = O(\sqrt{kd})$ for all $i \in [r]$,
    \item every FVS of $G$ with size at most $k$ contains some $X \in \mathcal{X}$ as a subset.
\end{enumerate}}

\medskip

Observe that this theorem directly gives us a subexponential FPT algorithm for FVS on induced-minor-free graphs with running time $2^{\widetilde{O}(k^{2/3})} n^{O(1)}$.
Indeed, we apply the STB theorem with $d = k^{1/3}$ to compute $\mathcal{X} = \{X_1,\dots,X_r\}$ in $2^{\widetilde{O}(k^{2/3})} n^{O(1)}$ time.
Then for each $X_i$, we compute a minimum FVS $S_i$ of $G-X_i$, which can be done in $2^{\widetilde{O}(\sqrt{kd})} n^{O(1)} = 2^{\widetilde{O}(k^{2/3})} n^{O(1)}$ time by standard DP on bounded-treewidth graphs.
By property (ii) above, there exists $X_i \in \mathcal{X}$ that is contained in some FVS of size at most $k$.
Thus, $X_i \cup S_i$ is also an FVS of size at most $k$, a solution we want.

To obtain the QPTAS is slightly more difficult.
The main idea is to properly combine the STB theorem with the bidimensionality technique~\cite{DemaineFominHajiaghayiThilikos2005,DemaineHajiaghayi2008,FominLokshtanovRamanSaurabh2011,FominLokshtanovSaurabhThilikos2020}.
First, we apply the STB theorem with $d = k/\log^c k$ for a large constant $c$, which takes quasi-polynomial time.
Let $\mathcal{X} = \{X_1,\dots,X_r\}$ be the collection returned and write $G_i = G-X_i$.
For each $i \in [r]$, $\mathbf{tw}(G_i) = k/\log^{c/2} k$.
Thus, there exists a balanced separator $Y_i \subseteq V(G_i)$ in the sense that every connected component of $G_i - Y_i$ has minimum FVS size at most $\frac{k}{2}$.
We then recursively apply the entire algorithm on each component $C$ of $G_i - Y_i$ to obtain an FVS $S_C$ of $C$.
The union of $S_C$'s and $Y_i$ is an FVS $S_i$ of $G_i$.
We do this for all $i \in [r]$ and return the set $X_i \cup S_i$ of the minimum size as the output.
The recursion terminates when the treewidth of the current graph is bounded by $f(\frac{1}{\varepsilon})$ for some function $f$; in this case, we apply the standard DP to compute a minimum FVS.
A careful analysis shows that choosing $f$ as an exponential function makes the returned FVS a $(1+\varepsilon)$-approximation solution, and the overall time complexity is quasi-polynomial.

\subsection{Proof sketch of the STB theorem} \label{sec:STB}
One of the key components of our proof for the STB theorem is a new structural result for $H$-induced-minor-free graphs (Corollary~\ref{cor:structure}), which is of independent interest.
This result bounds the treewidth of an $H$-induced-minor-free graph which admits a bounded-degeneracy FVS (or more generally, treewidth modulator) $M$ such that each connected component is neighboring to a few vertices in $M$.
Conventionally, for a set $V \subseteq V(G)$, we denote by $N_G(V)$ the set of neighbors of $V$ in $G$ that are not contained in $V$.
Below is the special case of the result for FVS.

\medskip

\noindent
\textbf{Structural result.}
{\it Let $G$ be an $H$-induced-minor-free graphs.
If there exists an FVS $M \subseteq V(G)$ of $G$ such that $G[M]$ is $\Delta$-degenerate and $|N_G(V(C))| \leq d$ for every connected component $C$ of $G-M$, then $\mathbf{tw}(G) = O(\sqrt{(d+\Delta) \cdot |M|})$}.

\begin{figure}[t]
\centering
\begin{tikzpicture}[x=0.94cm,y=0.94cm,font=\small,
  corevertex/.style={circle,draw=blue!65!black,fill=white,line width=.7pt,inner sep=2.2pt},
  component/.style={draw=teal!65!black,fill=teal!7,rounded corners=12pt,
    minimum width=2.4cm,minimum height=1.05cm,align=center,line width=.7pt},
  attachment/.style={draw=teal!65!black,line width=.8pt}]
\draw[red!60!black,fill=red!5,dashed,rounded corners=5pt]
  (0,0.2) rectangle (1.85,1.35);
\node[red!65!black] at (.925,1.05) {$X_i$};
\foreach \x in {.45,.925,1.4}{
  \draw[red!60!black,line width=.8pt] (\x-.07,.56)--(\x+.07,.70);
  \draw[red!60!black,line width=.8pt] (\x-.07,.70)--(\x+.07,.56);
}
\draw[blue!60!black,fill=blue!5,rounded corners=5pt,line width=.8pt]
  (2.5,0) rectangle (14,1.4);
\node[anchor=west,blue!65!black] at (2.7,1.02) {$M_i$};
\node[anchor=east] at (13.8,1.02)
  {$|M_i|=O(k)$,\quad $G[M_i]$ is $d$-degenerate};
\foreach \i/\x in {1/3.2,2/4.6,3/6.0,4/7.4,5/8.8,6/10.2,7/11.6,8/13.0}
  \coordinate (m\i) at (\x,.35);
\draw[blue!45!black] (m1)--(m2) (m3)--(m4) (m5)--(m6) (m7)--(m8);
\draw[blue!45!black] (m2) to[bend left=18] (m4);
\draw[blue!45!black] (m6) to[bend left=18] (m8);
\node[component] (c1) at (4.25,-2.0) {$C_1$\\a tree};
\node[component] (c2) at (8.25,-2.0) {$C_2$\\a tree};
\node[component] (c3) at (12.25,-2.0) {$C_3$\\a tree};
\draw[attachment] (m1)--(c1.north west) (m2)--(c1.north) (m3)--(c1.north east);
\draw[attachment] (m3)--(c2.north west) (m5)--(c2.north) (m6)--(c2.north east);
\draw[attachment] (m7)--(c3.north west) (m8)--(c3.north east);
\foreach \i in {1,...,8} \node[corevertex] at (m\i) {};
\node[fill=white,inner sep=3pt,text=teal!50!black] at (8.25,-.88)
  {$|N_F(C_j)\cap M_i|=O(d)$};
\node[align=center] at (8.25,-3.12)
  {Connected components of $G-(X_i \cup M_i)$: no edges between different components.};
\node[anchor=west,font=\small\bfseries] at (5,1.85) {The subgraph $G-X_i$};
\end{tikzpicture}
\caption{Illustrating the properties of the pairs $(X_i,M_i) \in \mathcal{Q}$.}
\label{fig:core-construction}
\end{figure}
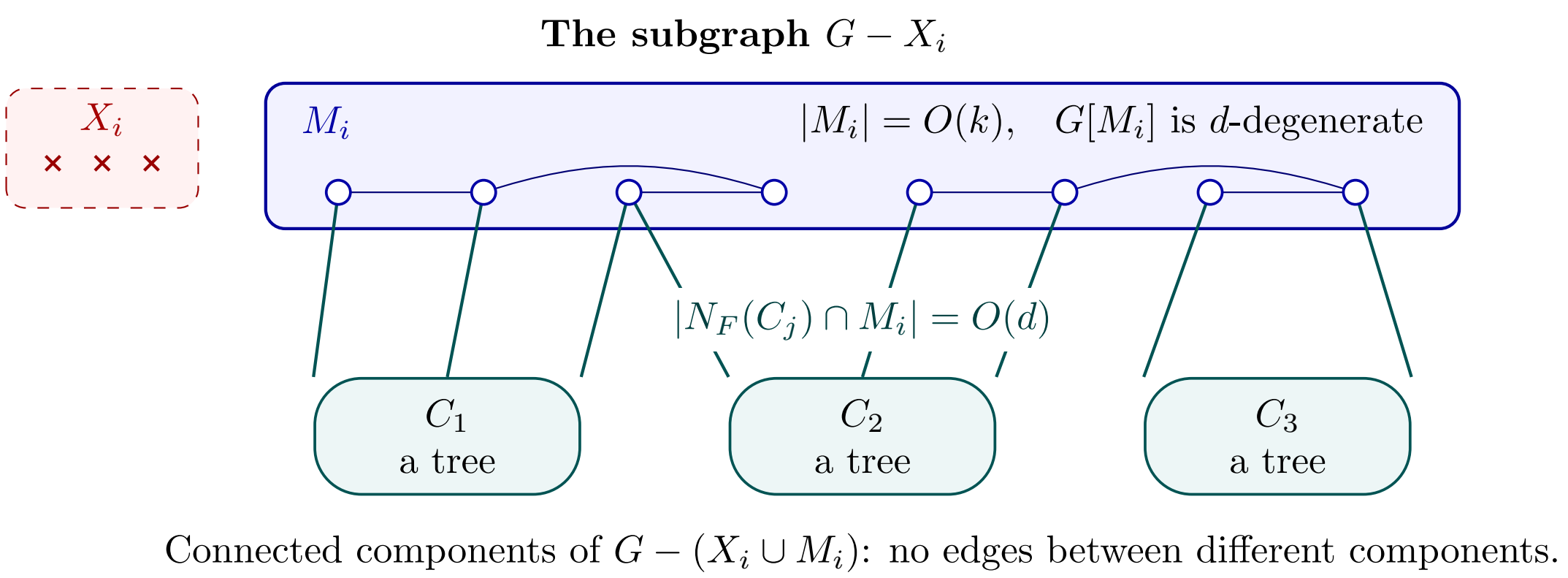

\medskip
Built on the above result, our branching algorithm basically works as follows.
Rather than just computing the collection $\mathcal{X}$ in the STB theorem, our branching procedure computes a collection $\mathcal{Q} = \{(X_1,M_1),\dots,(X_r,M_r)\}$ of pairs of subsets of $V(G)$, where $\mathcal{X} = \{X_1,\dots,X_r\}$ is the desired output, and each $M_i$ is a small FVS of $G - X_i$ that satisfies the nice conditions in the structural result and hence witnesses the low treewidth of $G-X_i$, as required in (i) of the theorem.
Formally, we require each $M_i$ to satisfy the following conditions:
\begin{enumerate}[(a)]
    \item $|M_i| = O(k)$,
    \item $G[M_i]$ is $d$-degenerate,
    \item $|N_{G-X_i}(V(C))| = O(d)$ for every connected component $C$ of $G-(X_i \cup M_i)$.
\end{enumerate}
See Figure~\ref{fig:core-construction} for an illustration.
The structural result then implies $\mathbf{tw}(G-X_i) = O(\sqrt{kd})$.
It turns out that condition (b) can be satisfied almost for free by adapting an algorithm of Korhonen and Lokshtanov~\cite{korhonen2024induced} (Lemma~\ref{lem:degenbranching}).
Hence, it suffices to make each $M_i$ satisfy conditions (a) and (c).

To obtain the collection $\mathcal{Q}$, we first compute a $2$-approximation FVS $M_\mathsf{init} \subseteq V(G)$.
We have $|M_\mathsf{init}| \leq 2k$.
In the branching procedure, we maintain two disjoint sets $X,M \subseteq V(G)$.
Initially, $X = \emptyset$ and $M = M_\mathsf{init}$.
At any point of the branching, we always have the invariant $X \cup M = M_\mathsf{init}$, which implies that $M$ is an FVS of $G-X$.
At (some of) the leaves of the branching, we add to $\mathcal{Q}$ the pair $(X_i,M_i)$ where $X_i = X$ and $M_i = M \cup L$ for a properly chosen set $L \subseteq V(G) \backslash (X \cup M)$ so that $M_i$ satisfies conditions (a) and (c).

Consider the current sets $X$ and $M$.
Let $F = G-(X \cup M)$, which is a forest.
For convenience, assume each tree of $F$ is rooted, and for a vertex $x \in V(F)$ contained in a tree $T$ of $F$, denote by $T_x$ the subtree of $T$ rooted at $x$.
We now greedily ``pack'' subtrees in $F$ which are neighboring to more than $d$ vertices in $M$.
Formally, starting with $\mathcal{T} = \emptyset$ and $F_0 = F$, we repeat the following procedure until no tree of $F_0$ has more than $d$ vertices in $M$:
\begin{enumerate}
    \item Find a tree $T$ of $F_0$ with $|N_G(V(T)) \cap M| > d$,
    \item Find a lowest node $x \in V(T)$ satisfying $|N_G(V(T_x)) \cap M| > d$,
    \item $\mathcal{T} \leftarrow \mathcal{T} \cup \{T_x\}$ and $F_0 \leftarrow F_0 - V(T_x)$.
\end{enumerate}
At the end, $\mathcal{T}$ consists of disjoint subtrees of trees in $F$ each of which has more than $d$ neighbors in $M$.
Let $R_\mathcal{T} \subseteq V(F)$ consist of the roots of the trees in $\mathcal{T}$.
The above packing procedure also guarantees the following good property: each connected component of $F - R_\mathcal{T}$ has at most $d$ neighbors in $M$.
Next, we distinguish two cases: $|\mathcal{T}| \leq 3k$ and $|\mathcal{T}| > 3k$.

If $|\mathcal{T}| \leq 3k$, then we add a pair $(X_i,M_i)$ to $\mathcal{Q}$ and terminate the current branch.
As mentioned before, we set $(X_i,M_i) = (X,M \cup L)$, and here $L$ is defined as the LCA closure of $R_\mathcal{T}$ in the trees in $F$.
We have $|L| \leq 2|R_\mathcal{T}| = 2|\mathcal{T}| = O(k)$.
Thus, condition (a) holds.
Each connected components of $G-(X_i \cup M_i) = F - L$ has at most $d$ neighbors in $M$ and at most two neighbors in $L$ (by the property of LCA closure), which implies $|N_{G-X_i}(V(C))| \leq d+2$ and hence condition (c) holds.

If $|\mathcal{T}| > 3k$, we first arbitrary remove some elements from $\mathcal{T}$ to make $|\mathcal{T}| = 3k$.
In this case, we shall create $O(k)$ branches as follows.
Let $S \subseteq V(G)$ be some unknown FVS of $G$ with $|S| \leq k$.
Then there are at least $2k$ trees in $\mathcal{T}$ that are disjoint from $S$, which we denote by $T_1,\dots,T_{2k}$ (these trees are unknown to us).
Each $T_i$ survives in $G-S$, and let $p_i$ be the number of neighbors of $T_i$ in $M \backslash S$.
Since $G-S$ is acyclic and $|M \backslash S| \leq |M| \leq 2k$, it turns out that $\sum_{i=1}^{2k} p_i \leq 4k-1$ and thus $p_j \leq 1$ for some $j \in [2k]$.
We then guess the tree $T_j$ in $\mathcal{T}$ and guess the (at most) one neighbor $v$ of $T_j$ in $M \backslash S$.
The number of guesses is $O(k)$, because $\mathcal{T} = O(k)$ and $|M| = O(k)$.
For each guess $(T_j,v)$, we move all neighbors of $T_j$ in $M$ except $v$ from $M$ to $X$, and proceed the branching procedure with the new $X$ and $M$.
Recall that $T_j$ has more than $d$ neighbors in $M$.
It follows that the new set $X$ (resp., $M$) contains at least $d$ more (resp., fewer) vertices than the old one, which is crucial when bounding the time complexity.
Furthermore, note that if the guess $(T_j,v)$ is correct, then all neighbors of $T_j$ in $M$ except $v$ must be contained in $S$; in this case, the vertices added to $X$ are all in $S$ and thus the corresponding branch is ``correct''.
Since a correct branch always exists, it turns out that our branching algorithm satisfies (ii) of the STB theorem.

To analyze the time complexity of the branching procedure, observe that the depth of the branching tree is $O(k/d)$.
Indeed, as observed above, going from one level to the next level, there are at least $d$ vertices moved from $M$ to $X$.
Since $|M_\mathsf{init}| \leq 2k$, there cannot be a branch with depth more than $2k/d$.
Since the degree of the branching tree is $O(k)$, the size of the branching tree is bounded by $2^{\widetilde{O}(k/d)}$, implying the $2^{\widetilde{O}(k/d)} n^{O(1)}$ running time.

\subsection{Proof sketch of the structural result}

Our structural result applied in the previous section is in turn the consequence of two structural lemmas (Lemma~\ref{lem:indeptorso} and Lemma~\ref{lem:sparsity}) for induced-minor-free graphs, which are also of independent interest and might find applications in other problems as well.

To present the two lemmas, we need to introduce the \textsc{Torso} operation on graphs.
Let $G$ be a graph and $M \subseteq V(G)$.
We define $\textsc{Torso}(G,M)$ as the graph obtained from $G$ by adding edges $(u,v)$ for all different $u,v \in M$ satisfying that there exists a connected component $C$ of $G-M$ with $u,v \in N_G(V(C))$.
In other words, $\textsc{Torso}(G,M)$ is obtained from $G$ by adding edges to make $N_G(V(C)) \cap M$ a clique for each connected component $C$ of $G-M$.

Our first structural lemma (Lemma~\ref{lem:indeptorso}) essentially states that applying \textsc{Torso} on the complement of an independent set preserves the induced-minor-freeness of the graph.

\medskip

\noindent
\textbf{Freeness-preserving lemma.}
{\it If $G$ is an $H$-induced-minor-free graph and $I$ is an independent set in $G$, then $\textsc{Torso}(G,V(G) \backslash I)$ is $H'$-induced-minor-free for some $H'$ depending on $H$.}

\smallskip

\begin{proof}[Proof sketch.]
Set $h = |V(H)|$.
Let $K_h^\bullet$ denote the subdivided $h$-clique, i.e., the graph obtained from $K_h$ by subdividing each edge once.
Without loss of generality, we may assume $H = K_h^\bullet$, because every $H$-induced-minor-free graph is also $K_h^\bullet$-induced-minor-free.
Let $G$ be a graph and $I$ be an independent set in $G$.
We shall show that if $G' = \textsc{Torso}(G,V(G) \backslash I)$ contains $H = K_h^\bullet$ as an induced minor, then $G$ also contains $H$ as an induced minor.

Write $X \subseteq V(H)$ as the set of the $h$ clique vertices of $H$ and $Y \subseteq V(H)$  as the set of the $\binom{h}{2}$ subdivision vertices of $H$.
Assume $G'$ contains $H$ as an induced minor.
For each $x \in V(H)$, let $G_x'$ be the connected subgraph of $G'$ corresponding to the vertex $x$ in the induced minor.
Each vertex $y \in Y$ has two neighbors $a_y,b_y$ in $H$, which are both vertices in $X$.
As such, we can choose $G_y'$ as an induced path in $G'$ with one endpoint adjacent to $G_{a_y}'$ and the other endpoint adjacent to $G_{b_y}'$, and except this, there is no adjacency between $G_y'$ and $G_x'$ for any $x \in V(H) \backslash \{y\}$.
To construct the $H$-induced-minor in $G$, we define
\begin{equation*}
    \textsc{Ext}(V) = V \cup \{v \in I: N_G[v] \cap V \neq \emptyset\},
\end{equation*}
for $V \subseteq V(G)$.
For each $x \in X$, define $P_x = \textsc{Ext}(V(G_x'))$.
Since $G_x'$ is connected, one can verify that $G[P_x]$ is also connected.
Then for each $y \in Y$, define $P_y = \textsc{Ext}(V(G_y')) \backslash (\bigcup_{x \in X} P_x)$.
Based on our choices of $G_y'$ for $y \in Y$, we manage to show that $G[P_y]$ is also connected for all $y \in Y$.
Furthermore, by carefully checking our construction of \textsc{Ext} (as well as the choices of $G_y'$) and exploiting the fact that $I$ is an independent set in $G$, one can verify that
\begin{enumerate}[(i)]
    \item $P_x \cap P_y = \emptyset$ for all different $x,y \in V(H)$,
    \item $P_x$ and $P_y$ are adjacent in $G$ iff $(x,y) \in E(H)$.
\end{enumerate}
Therefore, $G$ contains $H$ as an induced minor, in which each vertex $x \in V(H)$ corresponds to the (connected) subgraph $G[P_x]$.
\end{proof}

Our second structural lemma (Lemma~\ref{lem:sparsity}) bounds the increase of the number of edges after applying \textsc{Torso} to a ``nice'' vertex set of an induced-minor-free graph.

\medskip

\noindent
\textbf{Torso-sparsity lemma.}
{\it Let $G$ be an $H$-induced-minor-free graph, and $M \subseteq V(G)$ such that $G[M]$ is $\Delta$-degenerate and $|N_G(V(C))| \leq d$ for every connected component $C$ of $G-M$.
Then
\begin{equation*}
    |E(\textsc{Torso}(G,M))| - |E(G)| = O((d+\Delta) \cdot |M|).
\end{equation*}
}


\begin{proof}[Proof sketch.]
Write $E^* = E(\textsc{Torso}(G,M)) \backslash E(G)$ and we want to show $|E^*| = O((d+\Delta) \cdot |M|)$.
We shall construct a random $E_0 \subseteq E^*$ such that $\mathbb{E}[|E_0|] = O(|M|/(d+\Delta))$ and $\Pr[e \in E_0] = \Omega(\frac{1}{(d+\Delta)^2})$ for each edge $e \in E^*$, which will imply $|E^*| = O((d+\Delta) \cdot |M|)$.

Since $G[M]$ is $\Delta$-degenerate, we fix an orientation of the edges of $G[M]$ for which the out-degree of every vertex is at most $\Delta$.
Denote by $\varGamma(v)$ the set of out-neighbors of $v \in M$ in $G[M]$.
We first independently sample each vertex $v \in M$ with probability $p = \frac{1}{d+\Delta}$, and let $R \subseteq M$ consist of the sampled vertices.
Take the subset $I \subseteq R$ that consists of all $v \in R$ such that $\varGamma(v) \cap R = \emptyset$ (i.e., no out-neighbor of $v$ is sampled).
One can verify that $I$ is an independent set in $G[M]$ (and thus in $G$), and $\Pr[v \in I] \leq p(1-p)^\Delta = \Theta(p)$ for all $v \in M$.
The latter implies $\mathbb{E}[|I|] = \Theta(p|M|)$.

Write $G' = G - (M \backslash I)$.
Let $G_I'$ be the graph with $V(G_I') = I$ in which two vertices $u,v \in I$ are connected by an edge if there exists a component $C$ of $G'-I$ such that $N_G(V(C)) \cap I= \{u,v\}$.
Then we define $E_0 = E(G_I')$.
Since $G$ is $H$-induced-minor-free, one can show that $G_I'$ must be $H$-minor-free and thus $|E_0| = O(|I|)$, which implies $\mathbb{E}[|E_0|] = O(p|M|)$.

Consider an edge $e = (u,v) \in E^*$.
Since $e \in E(\textsc{Torso}(G,M))$ and $e \notin E(G)$, there exists a connected component $C$ of $G-M$ such that $u,v \in N_G(V(C))$.
By assumption, $|N_G(V(C))| \leq d$.
Note that $e \in E_0$ if both of the following events happen:
\begin{enumerate}[(i)]
    \item $u,v \in R$,
    \item $z \notin R$ for all $z \in \varGamma(u) \cup \varGamma(v) \cup (N_G(V(C)) \backslash \{u,v\})$.
\end{enumerate}
Since $|\varGamma(u) \cup \varGamma(v)| \leq 2 \Delta$ and $|N_G(V(C))| \leq d$, the probability that both events happen is at least $p^2 (1-p)^{(d+2\Delta)} = \Omega(p^2)$, which implies $\Pr[e \in E_0] = \Omega(p^2)$.
Thus, $\sum_{e \in E^*} \Pr[e \in E_0] = \Omega(p^2 |E^*|)$.
On the other hand, $\sum_{e \in E^*} \Pr[e \in E_0] = \mathbb{E}[|E_0|] = O(p|M|)$.
It follows that $|M| = \Omega(p|E^*|)$ and therefore $|E^*| = O(|M|/p) = O((d+\Delta) \cdot |M|)$.
\end{proof}

Now we combine the two lemmas above to prove the structural result in Section~\ref{sec:STB}.
Let $G$ be an $H$-induced-minor-free graphs and $M \subseteq V(G)$ be an FVS of $G$ such that $G[M]$ is $\Delta$-degenerate and $|N_G(V(C))| \leq d$ for every connected component $C$ of $G-M$.
Write $G^+ = \textsc{Torso}(G,M)$.
Applying the torso-sparsity lemma, we have
\begin{equation*}
    |E(G^+[M])| - |E(G[M])| \leq |E(G^+)| - |E(G)| = O((d+\Delta) \cdot |M|).
\end{equation*}
Note that $|E(G[M])| = O(\Delta|M|)$ since $G[M]$ is $\Delta$-degenerate.
Thus, $|E(G^+[M])| = O((d+\Delta) \cdot |M|)$.

Next, we show that $G^+[M]$ is $H'$-induced-minor-free for some $H'$ depending on $H$.
Let $G'$ be the graph obtained from $G$ by contracting each connected component of $G-M$ to a single vertex.
Then $G'$ is $H$-induced-minor-free.
Note that $I = V(G') \backslash M$ is an independent set in $G'$.
Therefore, the freeness-preserving lemma implies that $\textsc{Torso}(G',M)$ is $H'$-induced-minor-free for some $H'$.
Since $\textsc{Torso}(G',M) - I = G^+[M]$, $G^+[M]$ is $H'$-induced-minor-free as well.

Korhonen and Lokshtanov~\cite{korhonen2024induced} proved that $\mathbf{tw}(A) = O(\sqrt{|E(A)|})$ for an induced-minor-free graph $A$.
Therefore, the above discussion implies $\mathbf{tw}(G^+[M]) = O(\sqrt{(d+\Delta) \cdot |M|})$.
To further bound $\mathbf{tw}(G)$, we only need to apply the standard idea of gluing tree decompositions.
Let $\mathcal{T}$ be a tree decomposition of $G^+[M]$ with width $O(\sqrt{(d+\Delta) \cdot |M|})$.
Since $M$ is an FVS of $G$, each connected component $C$ of $G - M$ is a tree and thus admits a width-$1$ tree decomposition $\mathcal{T}_C$.
The neighbors of $C$ in $M$ form a clique in $G^+[M]$ due to the \textsc{Torso} operation, and hence there is one node $x$ of $\mathcal{T}$ whose bag contains all neighbors of $C$ in $M$.
We then append $\mathcal{T}_C$ to $x$ and add all vertices in the bag of $x$ to the bags of the nodes of $\mathcal{T}_C$.
After doing this for every component of $G - M$, we obtain a tree decomposition of $G$ with width $O(\sqrt{(d+\Delta) \cdot |M|})$.

\section{Preliminaries}

\paragraph{Basic notions.}
We write $\mathbb{N} = \{1,2,\dots\}$ as the set of natural numbers, and write $\mathbb{N}_0 = \mathbb{N} \cup \{0\}$.
For a number $n \in \mathbb{N}$, we write $[n] = \{1,\dots,n\}$ and $[n]_0 = \{0,1,\dots,n\}$.
The notation $\widetilde O(\cdot)$ denotes $O(\cdot)$ that suppresses logarithmic factors.

All graphs considered in this paper are finite, simple, and undirected.
Let $G$ be a graph.
The notation $V(G)$ denotes the set of vertices of $G$ and the notation $E(G)$ denotes the set of edges of $G$.
For $V \subseteq V(G)$, we use $N_G(V)$ to denote the set of vertices in $V(G) \backslash V$ that are neighboring to $V$, and define $N_G[V] = V \cup N_G(V)$.
Also, $G[V]$ denotes the subgraph of $G$ induced by $V$.

A graph is \textit{$d$-degenerate} if every nonempty subgraph has a vertex of degree at most $d$, or equivalently, it has an edge orientation in which every vertex has at most $d$ out-neighbors.
A graph property is \textit{hereditary} if it is preserved under taking induced subgraphs.

For a number $d \in \mathbb{N}$, a $d$-core of a graph $G$ is a subset $A \subseteq V(G)$ such that for every connected component $C$ of $G - A$, we have $|N_G(V(C))| \leq d$.
\begin{lemma} \label{lem:core1}
    Let $G$ be a graph and $V \subseteq V(G)$ be a subset.
    If $A \subseteq V(G)$ is a $d$-core of $G$, then $A \backslash V$ is a $d$-core of $G - V$.
\end{lemma}
\begin{proof}
Let $C$ be a connected component of $G - (V \cup A)$.
We need to show $|N_{G-V}(V(C))| \leq d$.
Note that $N_{G-V}(V(C)) \subseteq N_G(V(C)) \cap A$.
So it suffices to show $|N_G(V(C)) \cap A| \leq d$.
Since $A \subseteq V \cup A$, $C$ is contained in a connected component $C'$ of $G-A$.
We have $|N_G(V(C'))| \leq d$ because $A$ is a $d$-core of $G$.
As $V(C') \cap A = \emptyset$, it holds that $N_G(V(C)) \cap A \subseteq N_G(V(C'))$.
Thus, $|N_G(V(C)) \cap A| \leq |N_G(V(C'))| \leq d$.
\end{proof}

\begin{lemma} \label{lem:core2}
    Let $G$ be a graph.
    If $A \subseteq V(G)$ is a $d$-core of $G$ and $A' \subseteq V(G)$ is a $d'$-core of $G$, then $A \cup A'$ is a $(d+d')$-core of $G$.
\end{lemma}
\begin{proof}
Let $C$ be a connected component of $G - (A \cup A')$.
By Lemma~\ref{lem:core1}, $A$ is a $d$-core of $G-(A' \backslash A)$, which implies $|N_G(V(C)) \cap A| \leq d$.
For the same reason, $|N_G(V(C)) \cap A'| \leq d'$.
Therefore, we have $|N_G(V(C))| \leq |N_G(V(C)) \cap A|+|N_G(V(C)) \cap A'| \leq d+d'$.
\end{proof}

\paragraph{Tree decomposition and treewidth.}
In a rooted tree $T$, for a node $x \in V(T)$, we denote by $T_x$ the subtree of $T$ rooted at $x$.
Let $G$ be a graph.
A tree decomposition of $G$ is a pair $\mathcal{T}= (T,\beta)$ where $T$ is a rooted tree and $\beta:V(T) \rightarrow 2^{V(G)}$ is a map satisfying (i) $\bigcup_{x \in V(T)} \beta(x) = V(G)$, (ii) for every $(u,v) \in E(G)$, there exists $x \in V(T)$ with $u,v \in \beta(x)$, and (iii) for every $v \in V(G)$, $T[X_v]$ is connected where $X_v = \{x \in V(T): v \in \beta(x)\}$.
The \textit{width} of $\mathcal{T}$ is defined as $(\max_{x \in V(T)} |\beta(x)|) - 1$.
The \textit{treewidth} of $G$ is the minimum width of a tree decomposition of $G$.

\begin{lemma}\label{lem:treeDecomp}
    Given a connected graph $G$ with $n$ vertices, one can compute in $2^{O(\mathbf{tw}(G))} n$ time a tree decomposition $\mathcal{T}= (T,\beta)$ of $G$ with width $O(\mathbf{tw}(G))$ such that $G[\gamma(x)]$ is connected for all $x \in V(T)$, where $\gamma(x) = \bigcup_{y \in V(T_x)} \beta(y)$.
\end{lemma}
\begin{proof} Let $k=\mathbf{tw}(G)$. The case $n=1$ is immediate, so assume
$k\geq 1$. Korhonen's algorithm~\cite{korhonen2021treewidth}, given
an integer $q$, runs in time $2^{O(q)}n$ and either returns a tree
decomposition of width at most $2q+1$ or correctly reports that
$\mathbf{tw}(G)>q$. Trying $q=0,1,\ldots$ until a decomposition is
returned therefore gives, in total time $2^{O(k)}n$, a tree
decomposition of width $w\leq 2k+1=O(k)$.

We convert this decomposition into an elimination ordering
$v_1,\ldots,v_n$ of width at most $w$. Here, eliminating a vertex
means making its remaining neighbors pairwise adjacent and then
removing it; the width of the ordering is the maximum number of
remaining neighbors at any elimination step. Root the decomposition
arbitrarily. Repeatedly take a non-root leaf bag $B$ with parent
bag $B'$, eliminate the vertices of $B\setminus B'$ in any order,
and remove the leaf bag. Each such vertex occurs only in $B$ in
the current decomposition, so all its remaining neighbors lie in
$B$. Thus each elimination has at most $w$ remaining neighbors,
and every fill edge is covered by $B$. After these eliminations,
the surviving part of $B$ is contained in $B'$, so removing the
leaf bag preserves a tree decomposition of the remaining filled
graph. When only the root bag remains, eliminate its vertices
in any order.

Using a decomposition with $O(n)$ bags, this conversion and the
construction of the filled elimination graph take $O((w+1)^2n)$
time, which is absorbed by $2^{O(k)}n$.

We construct a rooted tree $T$ on $V(G)$ recursively. Take $v_n$ as
the root. For each connected component $D$ of $G-v_n$, choose as its
root the vertex with largest elimination index in $D$, attach this root
as a child of $v_n$, and apply the same construction to $G[D]$.
For each vertex $x=v_i$, let $C_x$ be the set consisting of $x$ and
its descendants in $T$. Equivalently, $C_x$ is the connected component
of $G[\{v_1,\ldots,v_i\}]$ containing $v_i$. In particular,
$G[C_x]$ is connected. Set
\[
    \beta(x)=\{x\}\cup N_G(C_x).
\]

We first bound the bag sizes. Every vertex $z\in N_G(C_x)$ has an
elimination index larger than $i$. As $G[C_x]$ is connected, there is
an $x$--$z$ path all of whose internal vertices have indices smaller
than $i$. After eliminating $v_1,\ldots,v_{i-1}$, the fill edges
therefore make $x$ adjacent to $z$. Hence $N_G(C_x)$ is contained in
the set of remaining neighbors of $x$ at its elimination step. We
conclude that $|\beta(x)|\leq w+1$.

We verify the tree-decomposition axioms. Each vertex $x$ belongs to
$\beta(x)$. For every edge $v_iv_j\in E(G)$ with $i<j$, we have
$v_j\in N_G(C_{v_i})$; thus both endpoints belong to $\beta(v_i)$.
Finally, fix $v\in V(G)$. If $v\in\beta(x)$ for $x\neq v$, then
$v\in N_G(C_x)$. The recursive construction implies that $v$ is an
ancestor of $x$. Furthermore, for any node $y$ on the path from $x$
to $v$ other than $v$, the containment $C_x\subseteq C_y$ gives
$v\in N_G(C_y)$, and hence $v\in\beta(y)$. Thus the bags containing
$v$ induce a connected subtree of $T$. This proves that $(T,\beta)$
is a tree decomposition of width at most $w$.

Let $\gamma(x)$ be the union of all bags at descendants of $x$,
including $x$ itself. We claim that
\[
    \gamma(x)=C_x\cup N_G(C_x).
\]
Indeed, every vertex of $C_x$ appears in its own bag, while
$N_G(C_x)\subseteq\beta(x)$. Conversely, if $y$ is a descendant of
$x$, then $C_y\subseteq C_x$, so
$\beta(y)\subseteq C_x\cup N_G(C_x)$. As $G[C_x]$ is connected and
every vertex of $N_G(C_x)$ has a neighbor in $C_x$, the induced graph
$G[\gamma(x)]$ is connected.

The tree $T$ and its bags can be recovered from the filled
elimination graph in $O(w^2n)$ time: the parent of each non-root
vertex is its smallest-indexed later neighbor, and the bag consists
of that vertex and its later neighbors. Thus the total running time
is $2^{O(k)}n$, as claimed.
\end{proof}

\begin{lemma}
\label{lem:balancesep}
    Let $G$ be a graph and $\mathcal{T} = (T,\beta)$ be a tree decomposition of $G$.
    For any subset $S \subseteq V(G)$ and any number $c \in \mathbb{N}$, there exist $r = O(c)$ nodes $x_1,\dots,x_r \in V(T)$ such that every connected component of $G - \bigcup_{i=1}^r \beta(x_i)$ contains at most $\frac{|S|}{c}$ vertices in $S$.
\end{lemma}

\begin{proof}
If $S=\emptyset$, take $r=0$. Otherwise, let $q=|S|/c$.
For each vertex $v\in S$, choose a node $a(v)\in V(T)$ such that
$v\in\beta(a(v))$. Assign each node $x\in V(T)$ the weight $w(x)=|\{v\in S:a(v)=x\}|$.
Thus, the total weight of $T$ is $|S|$.

We first construct a set $X\subseteq V(T)$ of fewer than $c$ nodes
such that every connected component of $T-X$ has weight at most $q$.
Root $T$ arbitrarily, and maintain a rooted tree $T'$, initially
equal to $T$. While the total weight of $T'$ exceeds $q$, choose
a deepest node $x$ whose subtree in $T'$ has weight greater than $q$.
Add $x$ to $X$, and remove its entire subtree from $T'$.
By the choice of $x$, each child subtree of $x$ in the current tree
has weight at most $q$.

The subtrees removed in different iterations are pairwise disjoint,
and each has weight greater than $q$. Consequently, writing
$r=|X|$, we obtain $rq<|S|$,
and hence $r<c$. Moreover, every connected component of $T-X$
is either a child subtree separated off in one of these iterations
or the tree remaining at termination. Each such component has
weight at most $q$.

Set $U=\bigcup_{x\in X}\beta(x)$.
For each vertex $v\in V(G)\backslash U$, the nodes whose bags contain
$v$ induce a connected subtree of $T$ disjoint from $X$, and
therefore lie in a single connected component of $T-X$.
If two vertices $u,v\notin U$ are adjacent in $G$, some bag contains
both of them, so their corresponding subtrees lie in the same
connected component of $T-X$. Thus, for every connected component
$C$ of $G-U$, there is a connected component $D$ of $T-X$ containing
every node whose bag contains a vertex of $C$.

In particular, $a(v)\in V(D)$ for every $v\in S\cap V(C)$. Hence $|S\cap V(C)| \le \sum_{x\in V(D)}w(x) \le q =\frac{|S|}{c}$.
Enumerating $X$ as $x_1,\ldots,x_r$ proves the lemma.
\end{proof}


\begin{lemma}\label{lem:findVZ}
Let $G$ be a graph with $n$ vertices, $M \subseteq V(G)$ satisfying $\mathbf{tw}(G-M) \leq t$, and $d \in \mathbb{N}$ be a parameter.
If $G-M$ contains a connected component $C$ with $|N_G(V(C)) \cap M| > d$, then one can compute in $2^{O(t)} n^{O(1)}$ time $V,Z \subseteq V(G) \backslash M$ satisfying the following conditions:
\begin{enumerate}[(i)]
    \item $Z \subseteq V$ and $N_{G-M}(V \backslash Z) \subseteq Z$,
    \item $G[V]$ is connected and $|N_G(V) \cap M| > d$,
    \item $|Z| = O(t)$,
    \item $M$ is a $d$-core of $G[M \cup V]-Z$.
\end{enumerate}
\end{lemma}

\begin{proof}
Let $C$ be a connected component of $G-M$ with $|N_G(V(C))\cap M|>d$. Apply Lemma~\ref{lem:treeDecomp} to $C$ to obtain a
rooted tree decomposition $(T,\beta)$ of width $O(t)$.
Thus, for each node $x$, 
$G[\gamma(x)]$ is connected.

Traverse $T$ in postorder. At each node $x$, order its
children arbitrarily as $y_1,\ldots,y_k$, and consider
successively the sets $W_j=\beta(x)\cup\bigcup_{i=1}^{j}\gamma(y_i)$ for $j=0,\ldots,k$.
Stop at the first set $W_j$ such that some connected
component of $G[W_j]$ has more than $d$ neighbors in $M$.
Let $V$ be the vertex set of this component, and set
$Z=V\cap\beta(x)$. The procedure must stop, because the
last set considered at the root is $V(C)$.

By construction, $Z\subseteq V\subseteq V(G)\backslash M$,
$G[V]$ is connected, and $|N_G(V)\cap M|>d$.
Furthermore, $|Z|\leq|\beta(x)|=O(t)$.
This proves~(ii) and~(iii).

For~(i), the tree-decomposition axioms imply that the
sets $\gamma(y_i)\backslash\beta(x)$ are pairwise disjoint
and satisfy $N_C\bigl(\gamma(y_i)\backslash\beta(x)\bigr) \subseteq\beta(x)$.
Every vertex of $V\backslash Z$ belongs to one of these
sets with $i\leq j$. Thus, all its neighbors in $G-M$
belong to $W_j$. Since $V$ is the vertex set of a
connected component of $G[W_j]$, these neighbors
belong to $V$.  Consequently, $N_{G-M}(V\backslash Z)\subseteq Z$.

To prove~(iv), observe that each child $y_i$ was fully
processed before $x$, including the set $\gamma(y_i)$.
Since the procedure did not stop earlier and
$G[\gamma(y_i)]$ is connected, $|N_G(\gamma(y_i))\cap M|\leq d$.
There are no edges between distinct sets
$\gamma(y_i)\backslash\beta(x)$. Hence every connected
component $D$ of $G[V\backslash Z]$ is contained in
$\gamma(y_i)\backslash\beta(x)$ for some $i\leq j$.
It follows that $|N_G(V(D))\cap M| \leq |N_G(\gamma(y_i))\cap M| \leq d$.
Therefore, $M$ is a $d$-core of $G[M\cup V]-Z$.

Finally, computing the decomposition takes
$2^{O(t)}n$ time. The total number of sets $W_j$
considered is $O(|V(T)|)$, since we consider only
prefixes of the ordered children at each node.
Computing their connected components and neighborhoods
takes polynomial time per set. Thus, the total running
time is $2^{O(t)}n^{O(1)}$.
\end{proof}

\begin{corollary} \label{cor:decomp}
Let $G$ be a graph with $n$ vertices, $M \subseteq V(G)$ satisfying $\mathbf{tw}(G-M) \leq t$, and $d \in \mathbb{N}$ be a parameter.
Then one can compute in $2^{O(t)} n^{O(1)}$ time a collection $\mathcal{P} = \{(V_1,Z_1),\dots,(V_p,Z_p)\}$ where $V_1,\dots,V_p,Z_1,\dots,Z_p \subseteq V(G) \backslash M$ satisfy the following conditions:
\begin{enumerate}[(i)]
    \item $V_1,\dots,V_p$ are disjoint and $Z_i \subseteq V_i$ for all $i \in [p]$,
    \item $G[V_i]$ is connected and $|N_G(V_i) \cap M| > d$ for all $i \in [p]$,
    \item $|Z_i| = O(t)$ for all $i \in [p]$,
    \item $M$ is a $d$-core of $G - (\bigcup_{i=1}^p Z_i)$.
\end{enumerate}
\end{corollary}

\begin{proof}
We give a recursive algorithm, proving its correctness by
induction on $|V(G)\backslash M|$. If $M$ is already a
$d$-core of $G$, return the empty collection, which satisfies
all the required conditions.

Otherwise, apply Lemma~\ref{lem:findVZ} to obtain sets $V_1,Z_1$. Since
$V_1\subseteq V(G)\backslash M$ is nonempty, the graph
$H:=G-V_1$ satisfies $M\subseteq V(H)$,
$\operatorname{tw}(H-M)\leq t$, and
$|V(H)\backslash M|<|V(G)\backslash M|$.
Recursively apply the algorithm to $H$ with the same set
$M$ and parameters $d,t$, obtaining a collection
\[
\mathcal{P}'=\{(V_2,Z_2),\ldots,(V_p,Z_p)\}.
\]
Return $\mathcal{P}:=\{(V_1,Z_1)\}\cup\mathcal{P}'$.

Conditions~(i)--(iii) follow from Lemma~\ref{lem:findVZ} and the induction
hypothesis. Indeed, all sets returned by the recursive call
are disjoint from $V_1$. Moreover, since $H$ is an induced
subgraph of $G$ containing $M$, for every $i\geq 2$ we have
$H[V_i]=G[V_i]$ and
$N_H(V_i)\cap M=N_G(V_i)\cap M$.

It remains to prove~(iv). Let
\[
Z^*:=\bigcup_{i=1}^{p}Z_i
\qquad\text{and}\qquad
H':=H-\bigcup_{i=2}^{p}Z_i.
\]
By the induction hypothesis, $M$ is a $d$-core of $H'$.
Furthermore, Lemma~4 gives
\[
N_{G-M}(V_1\backslash Z_1)\subseteq Z_1.
\]
Consequently, there are no edges between
$V_1\backslash Z_1$ and $V(H')\backslash M$.
These two sets partition $V(G-Z^*)\backslash M$.

Thus, every connected component $D$ of $(G-Z^*)-M$ is
either a connected component of $G[V_1\backslash Z_1]$
or a connected component of $H'-M$. In the first case,
$D$ has at most $d$ neighbors in $M$ because Lemma~4
ensures that $M$ is a $d$-core of $G[M\cup V_1]-Z_1$.
In the second case, the same bound follows from the
induction hypothesis. Since all vertices of $M$ are
retained, their adjacencies to $D$ are unchanged.
Therefore, $M$ is a $d$-core of $G-Z^*$, proving~(iv).

Finally, each recursive step removes at least one vertex,
so there are at most $n$ applications of Lemma~4. Each
application takes $2^{O(t)}n^{O(1)}$ time, and checking
whether $M$ is a $d$-core takes polynomial time.
The total running time is therefore
$2^{O(t)}n^{O(1)}$.
\end{proof}

\begin{sloppypar}
The following lemma uses the standard LCA-closure argument; see Bodlaender et al.~\cite[Lemma~5.4]{BodlaenderFLPST2016}
for a similar statement. We include a proof for completeness.
\end{sloppypar}

\begin{lemma} \label{lem:closure}
Let $G$ be a graph with $n$ vertices.
Given a subset $Z \subseteq V(G)$, one can compute in $2^{O(\mathbf{tw}(G))} n^{O(1)}$ time an $O(\mathbf{tw}(G))$-core $L \subseteq V(G)$ of $G$ such that $Z \subseteq L$ and $|L| = O(\mathbf{tw}(G)\cdot |Z|)$.
\end{lemma}
\begin{proof}
If $Z=\emptyset$, take $L=\emptyset$. Otherwise, put $k=\mathbf{tw}(G)$.
Using Korhonen's algorithm~\cite{korhonen2021treewidth}, trying successive
values of the width parameter as above, we compute a tree decomposition
$(T,\beta)$ of $G$ of width $w\leq 2k+1$ in time $2^{O(k)}n^{O(1)}$.
Root $T$ arbitrarily. For every $z\in Z$, choose a node $a_z$ with
$z\in\beta(a_z)$, and let $A=\{a_z:z\in Z\}$.
Let $Q$ consist of the root together with the closure of $A$ under
lowest common ancestors (LCAs).
Then $|Q|\leq 2|A|\leq 2|Z|$: in the minimal subtree containing the root
and $A$, every node added by LCA closure that is not already in $A$ is
a branching node, and there are at most $|A|-1$ such nodes; adding the
root contributes at most one further node.
Define
\[
L=\bigcup_{q\in Q}\beta(q).
\]
Clearly, $Z\subseteq L$ and $|L|\leq 2(w+1)|Z|=O((k+1)|Z|)$.

Every connected component $R$ of $T-Q$ has at most two neighbors in $Q$.
Indeed, it has at most one neighbor towards the root. If it had two
neighbors below it, their LCA would lie in $R$, contradicting the
LCA closure of $Q$.
Now let $C$ be a connected component of $G-L$.
For each vertex of $C$, the nodes whose bags contain it form a connected
subtree disjoint from $Q$. Since adjacent vertices occur together in a
bag and $C$ is connected, the union of these subtrees is connected.
Thus all bags meeting $V(C)$ lie in a single component $R$ of $T-Q$.

We claim that
\[
N_G(V(C))\subseteq\bigcup_{q\in N_T(V(R))}\beta(q).
\]
To see this, let $v\in N_G(V(C))$ and choose a neighbor $u\in V(C)$.
Some bag contains both $u$ and $v$, and its node lies in $R$.
Also, $v\in L$, so $v$ belongs to a bag at a node of $Q$.
The path between these two nodes crosses a node
$q\in N_T(V(R))$; by the connectedness condition of a tree decomposition,
$v\in\beta(q)$. This proves the claim. Consequently,
$|N_G(V(C))|\leq 2(w+1)=O(k+1)$, so $L$ is an $O(k+1)$-core of $G$.
The LCA closure and the union of the selected bags can be computed in
polynomial time in the size of the decomposition, giving the stated
running time.
\end{proof}

\paragraph{Treewidth modulators.}
For a graph $G$ and a number $t \in \mathbb{N}$, we say a subset $S \subseteq V(G)$ is a \textit{treewidth-$t$ modulator} of $G$ if $\mathbf{tw}(G-S) \leq t$.

\begin{lemma}[\cite{DBLP:conf/soda/GuptaLLM019}] \label{lem:twmod}
    Given a graph $G$ with $n$ vertices and a number $t \in \mathbb{N}$, one can compute in $2^{t^{O(1)}} n^{O(1)}$ time a treewidth-$t$ modulator $S$ of $G$ satisfying $|S| = O(k \log t)$, where $k$ is the minimum size of a treewidth-$t$ modulator of $G$.
\end{lemma}

\paragraph{Torso operation.}
For a graph $G$ and a set $M \subseteq V(G)$, we define $\textsc{Torso}(G,M)$ as the graph obtained from $G$ by adding edges $(u,v)$ for all different $u,v \in M$ satisfying that there exists a connected component $C$ of $G-M$ with $u,v \in N_G(V(C))$.
In other words, $\textsc{Torso}(G,M)$ is obtained from $G$ by adding edges to make $N_G(V(C)) \cap M$ a clique for each connected component $C$ of $G-M$.
We have the following simple observation.

\begin{lemma} \label{lem:torsotw}
    Let $G$ be a graph and $G^+ = \textsc{Torso}(G,M)$ for some $M \subseteq V(G)$.
    Then
    \begin{equation*}
        \mathbf{tw}(G) \leq \mathbf{tw}(G^+[M]) + \mathbf{tw}(G-M) + 1.
    \end{equation*}
\end{lemma}
\begin{proof} Write $w=\mathbf{tw}(G^+[M])$ and
$t=\mathbf{tw}(G-M)$. If either $M$ or $V(G)\setminus M$ is empty,
the asserted inequality is immediate. Otherwise, take a tree
decomposition $(T,\beta)$ of $G^+[M]$ of width $w$.

For each connected component $C$ of $G-M$, put
$S_C=N_G(C)\cap M$. By the definition of the torso, $S_C$ is a
clique in $G^+[M]$. Hence there is a node $x_C\in V(T)$ such that
$S_C\subseteq\beta(x_C)$. Take a tree decomposition
$(T_C,\beta_C)$ of $G[C]$ of width at most $t$ and enlarge each of
its bags to $\beta_C(y)\cup S_C$. Finally, join an arbitrary node of
$T_C$ to $x_C$ by an edge. Do this independently for every component
$C$ of $G-M$.

The resulting tree and bags form a tree decomposition of $G$.
Edges within $M$ and within each $C$ remain covered; every edge
between $M$ and $C$ has its endpoints in an enlarged bag of $T_C$.
For a vertex of $M$, the bags containing it stay connected because
each attached decomposition containing the vertex is joined at a bag
that contains $S_C$. Connectivity for a vertex outside $M$ follows
from its original component decomposition. Every enlarged bag has
size at most $(t+1)+(w+1)$, while the original bags have size at
most $w+1$. Therefore the new decomposition has width at most
$w+t+1$, proving the lemma.
\end{proof}

\paragraph{Minor-free and induced-minor-free graphs.}
A graph $H$ is an \textit{minor} of $G$ if $H$ can be obtained from $G$ by deleting vertices, deleting edges, and contracting edges.
A graph $H$ is an \textit{induced minor} of $G$ if $H$ can be obtained from $G$ by deleting vertices (but not edges) and contracting edges.
A graph $G$ is \textit{$H$-minor-free} (resp., \textit{$H$-induced-minor-free}) if $H$ is not a minor (resp., an induced minor) of $G$.
We denote by $\mathcal{G}_H$ the class of all $H$-induced-minor-free graphs.

The following treewidth bound uses the repaired separator theorem of Davies et al.~\cite{DaviesRzazewskiSpaldingJamiesonWalczak2026} together with the separator-number characterization of treewidth due to Dvo\v{r}\'{a}k and Norin~\cite{DvorakNorin2019}. Indeed, a separator of an induced subgraph also separates every spanning subgraph of it, so the separator bound applies to all subgraphs as required by that characterization.

\begin{lemma}[\cite{korhonen2024induced,DaviesRzazewskiSpaldingJamiesonWalczak2026}] \label{lem:separator}
    Let $H$ be a fixed graph.
    For any graph $G \in \mathcal{G}_H$, $\mathbf{tw}(G) = O(\sqrt{|E(G)|})$.
\end{lemma}

We also need the following lemmas about induced-minor-free graphs in our proofs.
For $h \in \mathbb{N}$, let $K_h$ be the complete graph with $h$ vertices and $K_h^\bullet$ be the graph obtained from $K_h$ by subdividing every edge (i.e., breaking the edge into two by inserting an intermediate vertex).

\begin{lemma} \label{lem:subdivclique}
    Let $H$ be a graph and $h = |V(H)|$.
    Then $\mathcal{G}_H \subseteq \mathcal{G}_{K_h^\bullet}$.
\end{lemma}
\begin{proof}
It suffices to show that $K_h^\bullet$ contains $H$ as an induced minor.
Let $X \subseteq V(K_h^\bullet)$ consist of the $h$ original vertices of $K_h^\bullet$ and $Y \subseteq V(K_h^\bullet)$ consist of the $\binom{h}{2}$ subdivision vertices of $K_h^\bullet$.
Fix a bijection $f: X \rightarrow V(H)$.
We consider every vertex $y \in Y$.
There are two vertices $u,v \in X$ neighboring to $y$ in $K_h^\bullet$.
If $(f(u),f(v)) \notin E(H)$, we delete $y$ from $K_h^\bullet$.
Otherwise, we contract the edge $(u,y)$ in $K_h^\bullet$.
Clearly, the resulting graph is isomorphic to $H$.
Thus, $K_h^\bullet$ contains $H$ as an induced minor.
It follows that if a graph $G$ contains $K_h^\bullet$ as an induced minor, then it also contains $H$ as an induced minor.
\end{proof}

\begin{lemma} \label{lem:minorfree}
    Let $G \in \mathcal{G}_H$ for some graph $H$ and $I \subseteq V(G)$ be an independent set in $G$.
    Define $G_I$ as the graph with $V(G_I) = I$ in which two vertices $u,v \in I$ are connected by an edge if there exists a connected component $C$ of $G-I$ such that $N_G(V(C)) = \{u,v\}$.
    Then $G_I$ is $H$-minor-free.
\end{lemma}
\begin{proof}
 Suppose for a contradiction that $G_I$ contains $H$ as an
ordinary minor. By the branch-set characterization of graph minors,
there is a subgraph $F$ of $G_I$ such that $H$ can be obtained from
$F$ by contracting edges only (and suppressing parallel edges).

For each edge $uv\in E(F)$, choose a connected component $C_{uv}$
of $G-I$ with $N_G(C_{uv})=\{u,v\}$. Different edges of $F$ require
different components, since the neighborhood of each such component
is a fixed pair of vertices. Consider the induced subgraph
\[
    G'=G\Bigl[V(F)\cup
           \bigcup_{uv\in E(F)}V(C_{uv})\Bigr].
\]
Since $I$ is independent, there are no edges between distinct
vertices of $V(F)$. Also, distinct chosen components have no edges
between them, and each $C_{uv}$ has neighbors in $V(F)$ only at
$u$ and $v$.

For each $uv\in E(F)$, contract the connected subgraph $C_{uv}$
together with $u$ into the vertex $u$. These contractions are valid
because $C_{uv}$ is connected and adjacent to $u$; they introduce
the edge $uv$ and no other edge among the vertices of $V(F)$.
After doing this for every edge of $F$, the resulting graph is
precisely $F$ (up to parallel edges). Thus $F$ is an induced minor
of $G$. Contracting the edges that turn $F$ into $H$ shows that $H$
is an induced minor of $G$, contradicting $G\in\mathcal{G}_H$.
Hence $G_I$ is $H$-minor-free.
\end{proof}

\section{Sublinear-treewidth branching theorem}

The main goal of this section is to prove the following theorem, which serves as the technical core of both our subexponential FPT algorithm and QPTAS.
Roughly speaking, given an induced-minor-free graph $G$ with an unknown treewidth modulator $S \subseteq V(G)$ of size at most $k$, the theorem allows us to guess in $2^{\widetilde{O}(k/d)} \cdot |V(G)|^{O(1)}$ time a subset $X$ of $S$ satisfying $\mathbf{tw}(G-X) = O(\sqrt{kd})$, via a branching algorithm.
Setting $d = k^c$ for a constant $c \in (0,1)$, the algorithm runs in subexponential time in $k$ and the treewidth of $G-X$ is sublinear in $k$.

\begin{restatable}{theorem}{branching} \label{thm:branching}
Let $H$ be a fixed graph.
Given a graph $G \in \mathcal{G}_H$ with $n$ vertices and three parameters $k,d,t \in \mathbb{N}$ with $d > 2t$, one can compute in $k^{t^{O(1)} k/d} n^{O(1)}$ time a collection $\mathcal{X} = \{X_1, \dots, X_r\}$ of $r = k^{t^{O(1)} k/d}$ subsets of $V(G)$ satisfying the following conditions:
\begin{enumerate}[(i)]
    \item $\mathbf{tw}(G-X_i) = t^{O(1)}\sqrt{kd}$ for all $i \in [r]$,
    \item for every treewidth-$t$ modulator $S \subseteq V(G)$ of $G$ satisfying $|S| \leq k$, there exists an index $i \in [r]$ such that $X_i \subseteq S$.
\end{enumerate}
\end{restatable}

\input{algo}

\subsection{Structural results}

\input{torso}

\input{sparsity}

\subsection{Putting everything together}


With the help of the results obtained in the previous sections, we are now ready to complete the proof of Theorem~\ref{thm:branching}.

\branching*

\begin{proof}
As mentioned in Section~\ref{sec:algo}, our final algorithm simply applies the algorithm of Lemma~\ref{lem:branching} to compute $\mathcal{Q} = \{(X_1,M_1),\dots,(X_r,M_r)\}$ and returns the collection $\mathcal{X} = \{X_1,\dots,X_r\}$.
By Lemma~\ref{lem:branching}, the time complexity is $k^{t^{O(1)} k/d} n^{O(1)}$ and $r = k^{t^{O(1)} k/d}$.
Also, condition~(ii) of the theorem follows directly from (iv) of Lemma~\ref{lem:branching}.
Therefore, it suffices to prove condition (i), i.e., $\mathbf{tw}(G-X_i) = t^{O(1)}\sqrt{kd}$ for all $i \in [r]$.

Write $G_i = G - X_i$.
By (i) of Lemma~\ref{lem:branching}, $\mathbf{tw}(G_i - M_i) \leq t$ and $|M_i| = O(t^2 \log t \cdot k)$.
Also, $M_i$ is an $O(d+t)$-core of $G-X_i$ and $G_i[M_i]$ is $d$-degenerate, by (ii) and (iii) of Lemma~\ref{lem:branching}, respectively.
Therefore, Corollary~\ref{cor:structure} implies $\mathbf{tw}(G_i) = O(t+ \sqrt{(d+t) \cdot t^2 \log t \cdot k}) = t^{O(1)} \sqrt{kd}$.
\end{proof}

\section{Dynamic programming}
In this section, we consider \textsc{$(t,\phi)$-Deletion} parameterized by the treewidth $w = \mathbf{tw}(G)$. 
We introduce an FPT algorithm with running time $w^{O(w)} n^{O(1)}$ by standard dynamic programming on a tree decomposition.
Combining this algorithm with Theorem~\ref{thm:branching} can already result in the subexponential FPT algorithm in Theorem~\ref{thm:subexp}.
To obtain the QPTAS in Theorem~\ref{thm:qptas}, however, we need the algorithm to solve a slightly more general version of the problem, which asks for not only an optimal solution of the instance but also the entire DP table.

To present the more general problem, we need to introduce the notion of \textit{boundaried graphs} and \textit{MSO-types}.
For a set $L$, an \textit{$L$-boundaried graph} is a pair $(G,\eta)$ where $G$ is a graph and $\eta:L \rightarrow V(G)$ is an injective map.
We call $L$ the \textit{boundary label set} of $G$.
If $(G,\eta)$ and $(G',\eta')$ are two $L$-boundaried graphs, then we use $(G,\eta) \oplus_L (H,\tau)$ to denote the graph (without boundary) obtained by gluing $(G,\eta)$ and $(G',\eta')$ as follows.
We first take the disjoint union $G \sqcup G'$, and then for each $\ell \in L$, add an edge $e_\ell = (\eta(\ell),\eta'(\ell))$ to $G \sqcup G'$.
After this, we contract all edges $e_\ell$ for $\ell \in L$, and the resulting graph is $(G,\eta) \oplus_L (H,\tau)$. 

Let $\phi$ be an MSO formula.
We define an equivalence relation $\sim_\phi$ among boundaried graphs as follows: $(G,\eta) \sim_\phi (G',\eta')$ if the following conditions hold:
\begin{enumerate}[(i)]
    \item $(G,\eta)$ and $(G',\eta')$ have the same boundary label set $L$,
    \item for all $L$-boundaried graphs, $(G,\eta) \oplus_L (H,\tau)$ satisfies $\phi$ iff $(G',\eta') \oplus_L (H,\tau)$ satisfies $\phi$.
\end{enumerate}
It is easy to verify that $\sim_\phi$ is an equivalence relation.
The \textit{$\phi$-type} of a boundaried graph is the equivalence class of $\sim_\phi$ containing the graph.
For convenience, graphs without boundary are viewed as $\emptyset$-boundaried graphs, and therefore also have their $\phi$-types.

A \textit{$(t,\phi)$-table} of a graph $G$ is a collection $\mathcal{S}$ of subsets of $V(G)$ satisfying that
\begin{enumerate}[(i)]
    \item $\mathbf{tw}(G-S) \leq t$ and $G-S$ satisfies $\phi$ for all $S \in \mathcal{S}$,
    \item for every $R \subseteq V(G)$ such that $\mathbf{tw}(G-R) \leq t$ and $G-R$ satisfies $\phi$, there exists $S \in \mathcal{S}$ with $|S| \leq |R|$ satisfying that $G-S$ is of the same $\phi$-type as $G-R$.
\end{enumerate}

It is easy to see that if $\mathcal{S}$ is a $(t,\phi)$-table of a graph $G$, then the smallest set $S \in \mathcal{S}$ is an optimal solution for \textsc{$(t,\phi)$-Deletion} on $G$.
In the more general version of the problem, we require the algorithm th return a $(t,\phi)$-table of the input graph $G$.
In the following lemma, we show this can be done in $w^{O(w)} n^{O(1)}$ time.
While the proof is a bit involved, the basic idea is similar to that used in MSO-checking on bounded-treewidth graphs~\cite{Courcelle1990,lokshtanov2026fine}.

\begin{lemma} \label{lem:boundedtw}
    Let $t \in \mathbb{N}$ be a fixed number and $\phi$ be a fixed MSO formula.
    Given a graph $G$ with $n$ vertices, one can compute a $(t,\phi)$-table of $G$ in $w^{O(w)} n^{O(1)}$ time, where $w = \mathbf{tw}(G)$.
\end{lemma}

\begin{proof}
All constants in the proof may depend on $t$ and $\phi$. Let $\theta_t$ be a fixed MSO sentence expressing that a graph has treewidth at most $t$. Such a sentence exists because, for fixed $t$, this property is characterized by excluding finitely many minors~\cite{RobertsonSeymour2004}, and containment of a fixed minor is MSO-expressible~\cite{CourcelleEngelfriet2012}. We first show how to summarize a residual graph of treewidth at most $t$ by a small representative. We then use these representatives in a dynamic program that retains a cheapest witness for every relevant type.

\paragraph{Equivalence under extensions.}
We identify boundary labels with the vertices carrying them and suppress the boundary maps from the notation. For two $L$-boundaried graphs $H_1,H_2$, write $H_1\equiv_L H_2$ if the graphs induced by their boundaries coincide under the label correspondence and, for every $L$-boundaried graph $F$ (with no restriction on its treewidth) and each $\psi\in\{\theta_t,\phi\}$, the graph $H_1\oplus_L F$ satisfies $\psi$ if and only if $H_2\oplus_L F$ does. Thus $\equiv_L$ refines $\sim_\phi$. Taking $F$ to consist only of isolated boundary vertices shows that equivalent graphs themselves agree on both $\theta_t$ and $\phi$. When $L=\emptyset$, gluing is disjoint union, so equivalence still records the behavior under every disjoint-union extension.

This equivalence is preserved by gluing along the boundary, forgetting boundary labels, and introducing a new boundary vertex with a prescribed neighborhood in the old boundary. For gluing, combine the graph being glued with any subsequent extension and apply the defining equivalence. For forgetting labels, augment an extension along the smaller boundary with isolated vertices carrying the forgotten labels. For introducing a vertex $v$, take an extension along the enlarged boundary, add the prescribed edges from $v$ to the old boundary, and regard $v$ as internal to this extension. In each case this gives an extension along the old boundary, to which the equivalence applies. Moreover, corresponding outputs have identical labeled boundary graphs whenever the inputs do. In particular, replacing either input of a gluing operation by an equivalent graph preserves the equivalence of the result.

We use the following standard consequence of the MSO composition theorem and the finite-state form of Courcelle's theorem~\cite{Courcelle1990,CourcelleEngelfriet2012}. For a fixed number $s$ of boundary labels, there is a finite refinement of $\equiv_L$ whose classes can be computed on graphs of fixed treewidth. Such a refinement is given by MSO types of sufficiently large fixed quantifier rank, with the boundary vertices named. Here a type records which sentences up to that rank hold, including adjacency between named boundary vertices; there are only finitely many such types up to logical equivalence. The composition theorem ensures that equality of types implies agreement on $\theta_t$ and $\phi$ after every gluing.

These types also yield an effective choice of representatives. First relabel the boundary by $[s]$, so the preprocessing does not depend on the particular labels. Adding the $s$ boundary vertices to every bag of a width-$t$ decomposition gives a decomposition of width at most $t+s$ in which all boundary vertices are present throughout. On such decompositions, a finite tree automaton computes the type and checks $\theta_t$. By enumerating its reachable states and retaining a witnessing decomposition for each state, we can precompute a representative of every type realized by a graph of treewidth at most $t$. Retain only witnesses satisfying $\theta_t$, so the representatives also have treewidth at most $t$. The automaton is fixed, hence this preprocessing and the representative sizes depend only on $s,t,\phi$. Given a graph of treewidth at most $t$, compute its type, choose the corresponding representative, and restore the original boundary labels. This takes polynomial time. We use this construction only for $s\leq 2(t+1)$.

\paragraph{Small representatives for larger boundaries.}
We next show that any $L$-boundaried graph $H$ of treewidth at most $t$, where $b=|L|$, can be replaced in polynomial time by an equivalent $L$-boundaried graph $J$ with $\mathbf{tw}(J)\leq t$ and $|V(J)|\leq c(b+1)$, for a constant $c$ depending only on $t$ and $\phi$. The additive term allows a nonempty representative when the boundary is empty.

The empty graph requires no replacement. Otherwise, compute a tree decomposition of $H$ of width at most $t$ in polynomial time~\cite{Bodlaender1996}, and make it rooted and binary by duplicating bags as needed. Denote the resulting decomposition by $(T,\beta)$. Mark the root and, for each boundary vertex, one node whose bag contains it. Let $Q$ be the closure of these nodes under lowest common ancestors, and put $Y=\bigcup_{x\in Q}\beta(x)$. The subtree connecting the marked nodes has at most $b$ branching nodes, so $|Q|=O(b+1)$ and $|Y|=O(b+1)$. Moreover, $Y$ contains every boundary vertex.

Each component $K$ of $T-Q$ has at most two neighbors in $Q$. Indeed, it has at most one neighbor above it, and two distinct neighbors below it would have their lowest common ancestor in $K$, contradicting the closure of $Q$. Every component of $T-Q$ has a neighbor in the nonempty set $Q$, and each node of $Q$ has degree at most three because $T$ is binary. Thus there are $O(|Q|)=O(b+1)$ such components. For each such component, let $A_K=(\bigcup_{x\in V(K)}\beta(x))\setminus Y$ and $C_K=N_H(A_K)$.

The nonempty sets $A_K$ partition $V(H)\setminus Y$, and no edge joins two distinct such sets. This follows because the bags containing a vertex outside $Y$ form a connected subtree disjoint from $Q$, and the endpoints of every edge occur together in a bag. Also, $C_K$ is contained in the union of the bags at the neighbors of $K$ in $Q$. To see this, a vertex of $Y$ adjacent to $A_K$ occurs in a bag of $K$ and a bag of $Q$. The connected subtree of bags containing it must therefore meet one of those neighboring bags. Consequently, $C_K\subseteq Y$ and $|C_K|\leq 2(t+1)$.

Keep $H[Y]$ and replace each $H[A_K\cup C_K]$ with $A_K\neq\emptyset$, viewed as a $C_K$-boundaried graph, by a constant-size equivalent representative. Give different representatives disjoint sets of fresh internal vertices, and identify their boundary vertices with the corresponding vertices of $Y$. The resulting graph $J$ has $O(b+1)$ vertices and the same labeled boundary graph as $H$. Each replacement preserves the truth of $\theta_t$ and $\phi$ even after any extension along $L$: the rest of the graph, together with that extension, is a context attached along $C_K$. All original boundary vertices lie in $Y$, so no replacement removes one. Replacing the pieces one at a time therefore proves $H\equiv_L J$. Taking the extension consisting of isolated boundary vertices shows that $J\models\theta_t$, and hence $\mathbf{tw}(J)\leq t$. All pieces have treewidth at most $t$ and constant-size boundaries, so the construction takes polynomial time. In particular, we compute logical types only at these constant-size interfaces.

For a fixed label set $L$, there are at most $2^{O((b+1)\log(b+2))}$ encodings of the resulting representatives. Each has $O(b+1)$ vertices and $O(b+1)$ edges, since its treewidth is at most the fixed constant $t$. Fix an order on the boundary labels and number the internal vertices consecutively. The vertex count and edge list then give the asserted bound. We use these graph encodings as states; different encodings may describe equivalent graphs, and we do not need to test this equivalence.

\paragraph{Dynamic programming.}
Using the treewidth approximation algorithm of Korhonen~\cite{korhonen2021treewidth}, and trying increasing width bounds until it succeeds, compute a tree decomposition of $G$ of width $W=O(w+1)$ in $2^{O(w)}n^{O(1)}$ time. Convert it into a nice tree decomposition with empty root and leaf bags and polynomially many nodes. For each node $x$, let $B_x$ be its bag, let $V_x$ be the union of the bags in its subtree, and put $G_x=G[V_x]$.

A state at $x$ is a pair $(D,J)$, where $D\subseteq B_x$ is the set of deleted bag vertices and $J$ is a representative with boundary $B_x\setminus D$. An entry stores a cost $a$ and a witnessing set $S_x\subseteq V_x$ of size $a$. We require $S_x\cap B_x=D$, $\mathbf{tw}(G_x-S_x)\leq t$, and $G_x-S_x\equiv_{B_x\setminus D}J$. Whenever several candidates have the same state encoding, keep one of minimum cost together with its witness. Witnesses may instead be stored implicitly using backpointers.

The transitions are as follows.
\begin{itemize}
\item At an empty leaf, store the empty graph with $D=\emptyset$, cost zero, and empty witness.
\item At an introduce node $x$ with child $y$, where $B_x=B_y\cup\{v\}$, extend each child state in two ways. Deleting $v$ replaces $D$ by $D\cup\{v\}$, leaves $J$ unchanged, adds $v$ to the witness, and increases the cost by one. Retaining $v$ leaves $D$, the witness, and the cost unchanged, and adds a new boundary vertex $v$ to $J$, adjacent precisely to $N_G(v)\cap(B_y\setminus D)$. These are all its neighbors in the retained part of $G_y$, because the tree-decomposition axioms imply that an introduced vertex has no neighbor in $V_y\setminus B_y$.
\item At a forget node $x$ with child $y$, where $B_x=B_y\setminus\{v\}$, replace $D$ by $D\setminus\{v\}$. If $v$ was retained in the child state, forget its boundary label in $J$ but retain the vertex itself. The cost and witness are unchanged; in particular, a deleted vertex remains in the witness after it leaves the bag.
\item At a join node $x$ with children $y,z$, we have $B_x=B_y=B_z$. Combine every pair of child states with the same deleted set $D$. Glue their representatives along $B_x\setminus D$, using disjoint copies of their internal vertices, and take the union of their witnesses. If the child costs are $a_y,a_z$, the new cost is $a_y+a_z-|D|$. Indeed, $V_y\cap V_z=B_x$, so the deleted bag vertices are exactly the vertices counted twice. There are no edges between $V_y\setminus B_x$ and $V_z\setminus B_x$, so gluing also gives exactly the residual graph of the combined witness, up to equivalence.
\end{itemize}
After each transition, test whether the resulting graph has treewidth at most $t$. Discard the candidate if it does not. Otherwise, apply the replacement procedure with its current boundary and store the resulting state and cost, retaining a cheapest witness when encodings coincide. Partial graphs need not satisfy $\phi$, so we test that property only at the root.

\paragraph{Correctness and extraction of the table.}
We prove two statements simultaneously by induction over the decomposition. First, every stored entry has a witness satisfying the requirements above. Second, for every $Z\subseteq V_x$ with $\mathbf{tw}(G_x-Z)\leq t$, there is a stored state $(D,J)$ with $D=Z\cap B_x$, cost at most $|Z|$, and $G_x-Z\equiv_{B_x\setminus D}J$.

Both statements hold at an empty leaf. For the first statement, each transition constructs the specified union or extension of child witnesses. The corresponding operation on their residual graphs agrees, up to equivalence, with the operation on their representatives. The extension properties of $\equiv_L$ and the replacement procedure preserve this equivalence. Agreement on $\theta_t$ makes the treewidth test valid for the actual residual graph as well. Retaining a cheapest witness for an encoding preserves all these requirements.

For the second statement, restrict $Z$ to each child vertex set. The resulting child residual graphs are induced subgraphs of $G_x-Z$, so their treewidth is at most $t$. Apply the induction hypothesis to these restrictions and use the corresponding transition. At an introduce node choose whether to delete $v$ according to $Z$; at a forget node use the forget transition; at a join node the chosen child states have the same deleted boundary $Z\cap B_x$. The resulting graph is equivalent to $G_x-Z$, and its cost is at most $|Z|$. At a join, the cost bound follows from $|Z\cap V_y|+|Z\cap V_z|-|Z\cap B_x|=|Z|$. The candidate passes the treewidth test because equivalence preserves $\theta_t$. Replacement preserves its equivalence. If another candidate with the same encoding has lower cost, its representative is identical and its witness remains suitable. Thus every partial deletion set is represented by an equivalent state of no greater cost; the induction does not require a particular encoding to survive.

At the empty root $r$, output the collection $\mathcal{S}$ of witnesses of all stored states whose representatives satisfy $\phi$. The first induction statement gives $\mathbf{tw}(G-S)\leq t$ and $G-S\models\phi$ for every $S\in\mathcal{S}$. Conversely, take any $Z\subseteq V(G)$ with $\mathbf{tw}(G-Z)\leq t$ and $G-Z\models\phi$. The second statement gives a root state of cost at most $|Z|$ whose representative is equivalent to $G-Z$. This representative satisfies $\phi$, so its witness $S$ is output. Both $G-S$ and $G-Z$ are equivalent to that representative under $\equiv_\emptyset$, and hence have the same $\phi$-type. Therefore $\mathcal{S}$ satisfies both requirements of a $(t,\phi)$-table. We need not determine which accepted states have the same $\phi$-type, since the definition permits redundant witnesses. If no state is accepted, there is no feasible deletion set, and the empty collection is a valid table.

\paragraph{Running time.}
There are $2^{O((W+1)\log(W+2))}$ states per node, including the choice of $D$. Each transition handles graphs with $O(W+1)$ vertices. Testing treewidth against the fixed bound $t$, computing the needed decompositions, and replacing the pieces all take polynomial time. Enumerating all pairs of child states at a join only squares the number of states and preserves this asymptotic bound. At the root, representatives have constant size, so testing $\phi$ takes constant time per state. Keeping or recovering witnesses adds a polynomial factor. The total running time is $2^{O((w+1)\log(w+2))}n^{O(1)}$, which is $w^{O(w)}n^{O(1)}$ for $w\geq2$; bounded values of $w$ are handled in polynomial time.
\end{proof}

\begin{corollary} \label{cor:DP}
    Let $t \in \mathbb{N}$ be a fixed number and $\phi$ be a fixed MSO formula.
    Then \textsc{$(t,\phi)$-Deletion} admits an algorithm with running time $w^{O(w)} n^{O(1)}$, where $w = \mathbf{tw}(G)$.
\end{corollary}
\begin{proof}
We simply apply Lemma~\ref{lem:boundedtw} to compute a $(t,\phi)$-table $\mathcal{S}$ of $G$ in $w^{O(w)} n^{O(1)}$ time and then output the smallest set $S_0 \in \mathcal{S}$.
By property (i) of a $(t,\phi)$-table, $S_0$ is a feasible solution.
To see it is optimal, consider an optimal solution $S^* \subseteq V(G)$ for \textsc{$(t,\phi)$-Deletion} on $G$.
By property (ii) of $(t,\phi)$-table, there exists $S \in \mathcal{S}$ such that $|S| \leq |S^*|$.
Therefore, $|S_0| \leq |S| \leq |S^*|$.
\end{proof}

\section{Subexponential FPT algorithms}
\label{sec:subexp}

In this section, we combine Theorem~\ref{thm:branching} and Corollary~\ref{cor:DP} to prove Theorem~\ref{thm:subexp}, which gives subexponential FPT algorithms for \textsc{$(t,\phi)$-Deletion}.


\subexp*
\begin{proof}
We first apply Theorem~\ref{thm:branching} with $d = k^{1/3}$ to compute the collection $\mathcal{X}= \{X_1,\dots,X_r\}$ in $2^{\widetilde{O}(k^{2/3})} n^{O(1)}$ time, where $r = 2^{\widetilde{O}(k^{2/3})}$.
For each $i \in [r]$, we apply Corollary~\ref{cor:DP} on the graph $G-X_i$, which gives us a minimum subset $S_i \subseteq V(G) \backslash X_i$ such that $\mathbf{tw}(G-(X_i \cup S_i)) \leq t$ and $G-(X_i \cup S_i)$ satisfies $\phi$.
By (i) of Theorem~\ref{thm:branching}, $\mathbf{tw}(G-X_i) = O(\sqrt{kd}) = O(k^{2/3})$ for all $i \in [r]$.
Therefore, computing each $S_i$ can be done in $2^{\widetilde{O}(k^{2/3})} n^{O(1)}$ time.
Since $r = 2^{\widetilde{O}(k^{2/3})}$, the overall time cost is also $2^{\widetilde{O}(k^{2/3})} n^{O(1)}$.
Define $i^* = \arg\min_{i \in [r]} |X_i \cup S_i|$.
If $|X_{i^*} \cup S_{i^*}| \leq k$, then we return $X_{i^*} \cup S_{i^*}$ as the output, otherwise we return \textsc{Fail}.

To see the correctness of the algorithm, observe that if the algorithm returns a subset of $V(G)$, then it is a feasible solution for the instance.
So it suffices to show that if there exists a solution $S \subseteq V(G)$ for \textsc{$(t,\phi)$-Deletion} on $G$ with $|S| \leq k$, then the algorithm does not fail.
Note that $S$ is a treewidth-$t$ modulator of $G$.
Thus, by (ii) of Theorem~\ref{thm:branching}, there exists $i \in [r]$ such that $X_i \subseteq S$.
Now $S \backslash X_i$ is a feasible solution on $G-X_i$, and thus $|S_i| \leq |S \backslash X_i|$.
It follows that
\begin{equation*}
    |X_{i^*} \cup S_{i^*}| \leq |X_i \cup S_i| = |X_i| + |S_i| \leq |X_i| + |S \backslash X_i| \leq |S| \leq k.
\end{equation*}
Thus, the algorithm returns $X_{i^*} \cup S_{i^*}$ rather than \textsc{Fail}.
\end{proof}

\section{Quasipolynomial-time approximation schemes} \label{sec:QPTAS}

In this section, we use Theorem~\ref{thm:branching} to obtain a QPTAS for \textsc{$(t,\phi)$-Deletion}.
The approximation scheme is proved by repeatedly applying Theorem~\ref{thm:branching} to instances reduced using balance separator (Lemma~\ref{lem:balancesep}). 
For every graph $G'$,
 define
\[
 \operatorname{OPT}_{t,\phi}(G')
 =\min\{|S|:S\subseteq V(G'),\ \operatorname{tw}(G'-S)\leq t \textit{ and } G'-S\models\phi\}.
\]

Throughout this section, \(G\) denotes the input graph of \(\textsc{$(t,\phi)$-Deletion}\) where $G\in \mathcal{G}_H$, and we assume that \(k=\operatorname{OPT}_{t,\phi}(G)\) and this value is known to the algorithm. The high-level idea is to apply Theorem~\ref{thm:branching}, which guarantees that, for an optimum solution \(S^{\star}\), there exists a set \(X\) in the collection output by the algorithm such that \(G-X\) has treewidth sublinear in \(k\). We then aim to divide the problem into two subinstances, each with an optimum solution size smaller by a constant factor. To achieve this, we fix an approximate solution and apply Lemma~\ref{lem:balancesep} to find a separator that partitions the approximate solution in a balanced manner. We recursively solve the resulting subinstances and combine their $(t,\phi)$-tables 
to obtain $(t,\phi)$-table of the graph obtained after deleting \(X\) and the separator. 
We will show that 
the resulting 
$(t,\phi)$-table
and 
is enough to extract
a good approximate solution to the original instance.

Our algorithm is a branching algorithm and each local branching step is an application of Theorem~\ref{thm:branching} and Lemma~\ref{lem:balancesep}.

\begin{lemma}[Local branching]
\label{lem:qp-local}
Let $0<\delta\leq 1/2$.
There are constants $c_1,c_2,c_3>0$, depending only on $H,t$,
with the following property. 
For every sufficiently large $q$, where the threshold depends
only on $H,t,\delta$, given a graph $G'\in\mathcal G_H$
and a set $W\subseteq V(G')$ of size $q$, one can compute
a family $\mathcal Y$ 
of size
$2^{c_2\delta^{-3}\log^9q}$ in time 
$2^{c_2\delta^{-3}\log^9q}|V(G')|^{O(1)}$ and for each $Y\in\mathcal Y$,
a set $Z_Y\subseteq V(G')\backslash Y$ and two
vertex-disjoint induced subgraphs $G'_{Y,1},G'_{Y,2}$
such that:
\begin{enumerate}[(i)]
\item $G'-(Y\cup Z_Y)$ is the disjoint union of
$G'_{Y,1}$ and $G'_{Y,2}$;

\item $|W\cap V(G'_{Y,j})|\leq 2q/3$
for $j\in\{1,2\}$;

\item $|Z_Y|\leq c_1q/\log^3q$;

\item $\operatorname{tw}(G'_{Y,j})\leq
c_3q/\log^4q$ for $j\in\{1,2\}$;

\item For every treewidth-$t$ modulator
$S\subseteq V(G')$ satisfying $|S|\leq q/\delta$,
there exists some $Y\in\mathcal Y$ such that
$Y\subseteq S$.
\end{enumerate}
\end{lemma}

\begin{proof}

Set
\[
 b=\left\lceil \frac{q}{\delta}\right\rceil,
 \qquad
 d=\left\lceil \frac{\delta q}{\log^8 q}\right\rceil.
\]
Choose the threshold sufficiently large so that
$q\geq 1/\delta$ and
$\delta q/\log^8 q\geq 1$.
Then
\[
 b\leq \frac{2q}{\delta},
 \qquad
 \frac{\delta q}{\log^8 q}
 \leq d\leq\frac{2\delta q}{\log^8 q},
 \qquad
 \frac{b}{d}\leq
 \frac{2\log^8 q}{\delta^{2}}.
\]

Apply Theorem~\ref{thm:branching} to $G'$ with parameters $k=b,d$, and $t$.
Let $\mathcal Y$ denote the resulting family.
Its cardinality is at most
\[
 b^{O(t^2b/d)}
 \leq 2^{c_2\delta^{-3}\log^9 q},
\]
where the inequality follows from
$\log b \leq \log q \cdot \log (1/\delta)$, and $c_2$ is a sufficiently
large constant depending only on $H,t$.
The containment guarantee (condition $(ii)$) of Theorem~\ref{thm:branching}
implies property~$(v)$. Moreover, for every $Y\in\mathcal Y$, we have
\[
 \operatorname{tw}(G'-Y)
 \leq C_{H,t}\sqrt{bd}
 \leq 2C_{H,t}\frac{q}{\log^4 q},
\]
where $C_{H,t}$ is a constant. Thus, the condition $(iv)$ follows. 
We next construct the set $Z_Y$ and the two
subgraphs $G'_{Y,1}$ and $G'_{Y,2}$ for each
$Y\in\mathcal Y$. Fix $Y\in\mathcal Y$. By the polynomial-time
treewidth approximation algorithm of Feige,
Hajiaghayi, and Lee~\cite{FHL2008}, we can
compute a tree decomposition of $G'-Y$ of width
\(
 O\bigl( w\sqrt{\log w}\bigr),
\)
where $w=\operatorname{tw}(G'-Y)$. Thus, we can obtain a tree decomposition $\mathcal{T}=(T,\beta)$ of $G'-Y$
of width at most
\(
 O(\frac{q}{\log^3 q}).
\)
Now we apply Lemma~\ref{lem:balancesep} with $S=W$ and $c=2$ to get $O(1)$ nodes $x_1,\dots,x_r\in T$ such that each connected component of $G'-(Y\cup (\bigcup_{i=1}^r \beta(x)))$ has size at most $|W|/2$ vertices from $W$. We set $Z_Y=\bigcup_{i=1}^r \beta(x)$. Notice that \(
 |Z_Y|\leq c_1\frac{q}{\log^3 q}
\) where $c_1$ is a constant. 
This establishes
property~$(iii)$.

It remains to construct the two subgraphs.
Consider the connected components of
$G'-(Y\cup Z_Y)$, and assign to each
component $C$ the weight $|W\cap V(C)|$.
The total weight is at most $q$, and each
component has weight at most $q/2$.
If some component has weight in
$[q/3,2q/3]$, place it in the first group
and all remaining components in the second
group. Otherwise, every component has
weight less than $q/3$. If the total weight
is at least $q/3$, greedily add components
to the first group until its weight first
reaches $q/3$. The resulting weight is less
than $2q/3$, and the weight of the remaining
components is at most $2q/3$. If the total
weight is less than $q/3$, place all
components in the first group and leave
the second group empty. Components of
weight zero can be assigned arbitrarily.

Let $G'_{Y,1}$ and $G'_{Y,2}$ be the
subgraphs induced by the vertices in the
two groups, respectively. By construction,
they are vertex-disjoint, there are no
edges between them, and
\[
 G'-(Y\cup Z_Y)
 =G'_{Y,1}\mathbin{\dot\cup}G'_{Y,2}.
\]
Furthermore,
\[
 |W\cap V(G'_{Y,j})|
 \leq\frac{2q}{3}
 \qquad\text{for }j\in\{1,2\}.
\]
Hence properties~$(i)$ and~$(ii)$ hold.
All computations following the application
of Theorem~\ref{thm:branching} take polynomial time for
each $Y\in\mathcal Y$. So, the total
running time is at most
\(
 2^{c_2\delta^{-3}\log^9 q}
 |V(G')|^{O(1)},  
\) for some contant $c_2$. 
\end{proof}

Now we are ready to give the QPTAS. 

\qptas*

\begin{proof}
Fix $0<\varepsilon\leq 1$, and set $\delta=\varepsilon/2$.
All constants below may depend on $H,t,\phi,\varepsilon$.
We first describe the type information used by the algorithm.
Let $\mathcal T$ be the finite set of $\phi$-types of graphs
with empty boundary, and write $\operatorname{tp}(F)$ for
the type of a graph $F$.
Since the type equivalence is preserved by disjoint union,
there is a well-defined operation $\oplus$ on $\mathcal T$
such that
\[
 \operatorname{tp}(F_1\mathbin{\dot\cup}F_2)
 =
 \operatorname{tp}(F_1)\oplus\operatorname{tp}(F_2).
\]

Define a relation on $\mathcal T$ by putting
$\sigma\preceq_0\tau$ if there exist graphs $F_\sigma,F_\tau$
of types $\sigma,\tau$, respectively, such that $F_\sigma$
is an induced subgraph of $F_\tau$.
Let $\preceq$ be its reflexive transitive closure of $\preceq_0$.
In particular, deleting vertices can only decrease the type
with respect to $\preceq$.
The preorder $\preceq$ (i.e., the relation is reflexive and transitive) is compatible with disjoint union:
\[
 \sigma_1\preceq\tau_1,\quad
 \sigma_2\preceq\tau_2
 \quad\Longrightarrow\quad
 \sigma_1\oplus\sigma_2
 \preceq
 \tau_1\oplus\tau_2.
\]
Indeed, for a generating relation
$\sigma\preceq_0\tau$, take witnessing graphs
$F_\sigma\subseteq_{\mathrm{ind}}F_\tau$ and any graph $K$
of type $\kappa$.
Then
$F_\sigma\mathbin{\dot\cup}K$ is an induced subgraph of
$F_\tau\mathbin{\dot\cup}K$, proving
$\sigma\oplus\kappa\preceq_0\tau\oplus\kappa$.
The claim follows by transitivity and by applying this
observation to each coordinate.
For each $\tau\in\mathcal T$, let $\chi_\tau$ be the fixed
MSO property
\[
 F\models\chi_\tau
 \quad\Longleftrightarrow\quad
 \operatorname{tp}(F)\preceq\tau.
\]
This property is hereditary.
Moreover, if graphs of type $\tau$ satisfy $\phi$, then every
graph satisfying $\chi_\tau$ also satisfies $\phi$:
this follows from the hereditariness of $\phi$ along each
generating relation $\preceq_0$ of $\preceq$.
All type information, the operation $\oplus$, and the preorder
$\preceq$ are fixed independently of the input.

Let $G\in\mathcal G_H$ be the input graph, and put
\(
 n=|V(G)|\) and \(
 k=\operatorname{OPT}_{t,\phi}(G).
\)
Using Lemma~\ref{lem:twmod}, compute a treewidth-$t$
modulator $M$ whose size is at most a constant multiple of
$k$.
On the bounded-treewidth graph $G-M$, use
Lemma~\ref{lem:boundedtw} to compute a minimum deletion set
$P\subseteq V(G)\backslash M$ such that
$G-(M\cup P)\models\phi$.
If $S^\star$ is an optimum solution, then
$G-(M\cup S^\star)$ is an induced subgraph of
$G-S^\star$ and therefore satisfies $\phi$.
Consequently,
\[
 |P|\leq |S^\star\backslash M|\leq k.
\]
Let
\(
 W=M\cup P \text{ and } q=|W|\leq a_t k
\)
for a constant $a_t$. 
Notice that 
\(
 \operatorname{tw}(G-W)\leq t \text{ and }
 G-W\models\phi.
\)
We call the vertices of $W$ marked.
We maintain the marked set by restriction:
a recursive call has input
\[
 G'\subseteq_{\mathrm{ind}}G,
 \qquad
 W'=W\cap V(G'),
 \qquad
 q'=|W'|.
\]
In particular, $W'$ is a treewidth-$t$ modulator of $G'$.
The output of a call consists of at most one deletion set
$S_\tau$ for each $\tau\in\mathcal T$.
Every stored set must satisfy
\[
 \operatorname{tw}(G'-S_\tau)\leq t,
 \qquad
 \operatorname{tp}(G'-S_\tau)\preceq\tau.
\]
If several candidates are available for an entry, retain
one of minimum cardinality.
An entry is absent if no candidate has been obtained. Choose a sufficiently large integer $s_0\geq 2$, specified
below, and put $B=t+s_0$.
If $\operatorname{tw}(G')\leq B$, compute every entry exactly
using Lemma~\ref{lem:boundedtw}, applied to the fixed
hereditary properties $\chi_\tau$.
This takes polynomial time because $B$ is a constant.
Suppose now that $\operatorname{tw}(G')>B$.
Since
\(
 \operatorname{tw}(G')\leq t+|W'|,
\)
we have $q'>s_0$.
Choose $s_0$ large enough that
Lemma~\ref{lem:qp-local} is applicable with the fixed value
$\delta=\varepsilon/2$.

First, for every $\tau\in\mathcal T$, compute a minimum set
$P_\tau\subseteq V(G')\backslash W'$ such that
\(
 G'-(W'\cup P_\tau)\models\chi_\tau.
\)
This is a bounded-treewidth computation on $G'-W'$, using Lemma~\ref{lem:boundedtw}. 
Whenever such a set exists, include $W'\cup P_\tau$ as a
candidate for entry $\tau$.
Next, apply Lemma~\ref{lem:qp-local} to $(G',W')$.
For every returned set $Y$, let $Z_Y,G'_{Y,1},G'_{Y,2}$
be the associated separator and two induced subgraphs.
Recursively process
\[
 \bigl(G'_{Y,j},\,W'\cap V(G'_{Y,j})\bigr),
 \qquad j\in\{1,2\}.
\]
For every pair of available child entries
$S_{1,\tau_1},S_{2,\tau_2}$, form
\[
 D=Y\cup Z_Y\cup S_{1,\tau_1}\cup S_{2,\tau_2}.
\]
For every $\tau$ satisfying
$\tau_1\oplus\tau_2\preceq\tau$, include $D$ as a candidate
for entry $\tau$. These candidates are feasible.
Indeed,
\[
 G'-D
 =
 \bigl(G'_{Y,1}-S_{1,\tau_1}\bigr)
 \mathbin{\dot\cup}
 \bigl(G'_{Y,2}-S_{2,\tau_2}\bigr),
\]
so its treewidth is at most $t$.
Compatibility of $\preceq$ with disjoint union gives
\[
 \operatorname{tp}(G'-D)
 \preceq \tau_1\oplus\tau_2
 \preceq\tau.
\]

We next bound the separator cost.
Consider any binary recursion tree obtained by choosing
one set $Y$ at each internal call.
If a node has $m$ marked vertices and its children have
$m_1,m_2$ marked vertices, then
\(
 m_1+m_2\leq m \text{ and }
 m_j\leq \frac{2m}{3}.
\)
Every internal node has $m>s_0$, and its separator has
size at most
\(
 \frac{c_1m}{\log^3 m}.
\)
Charge its separator cost equally to its $m$ marked vertices.
Each marked vertex receives a charge of at most
$c_1/\log^3 m$ at that node.

Along a root-to-leaf path, the positive marked-set sizes
decrease by a factor of at least $3/2$.
Reading the internal nodes in the reverse order, their sizes
are therefore at least
$s_0,(3/2)s_0,(3/2)^2s_0,\ldots$.
Hence the total charge to any marked vertex is at most
\[
 c_1\sum_{i\geq 0}
 \frac{1}{\log^3(s_0(3/2)^i)}
 \leq
 \frac{C}{\log^2 s_0}
\]
for a constant $C$.
Thus, any such recursion tree starting with $m$ marked
vertices has total separator cost at most
\[
 \frac{C \cdot m}{\log^2 s_0}.
\]
Equivalently, the total separator cost can be bounded by analysing the recurrence in which a subproblem of size \(m\) incurs a separator cost of at most
$c_1m/\log^3 m$, child sizes satisfying the two
inequalities above, and zero cost at leaves. We prove the approximation guarantee by induction on the
recursion.
Fix a state $\tau$ and any set $R\subseteq V(G')$ satisfying
\[
 \operatorname{tw}(G'-R)\leq t,
 \qquad
 \operatorname{tp}(G'-R)\preceq\tau.
\]
We show that the algorithm obtains a feasible entry of cost
at most $(1+\delta)|R|$ plus the separator costs in a
suitable binary recursion tree.
At a bounded-treewidth leaf, the entry is computed exactly,
so its cost is at most $|R|$.

At an internal call, suppose first that $|R|>q'/\delta$.
The graph $G'-(W'\cup R)$ is an induced subgraph of
$G'-R$, and therefore satisfies $\chi_\tau$.
By the optimality of $P_\tau$,
\(
 |P_\tau|\leq |R\backslash W'|.
\)
The fallback candidate consequently satisfies
\[
 |W'\cup P_\tau|
 \leq q'+|R\backslash W'|
 \leq q'+|R|
 <(1+\delta)|R|.
\]
In this case, the comparison tree terminates at this call
and incurs no separator cost.
Otherwise, $|R|\leq q'/\delta$.
Since $R$ is a treewidth-$t$ modulator of $G'$,
Lemma~\ref{lem:qp-local} guarantees a branch with $Y\subseteq R$.
Fix this branch and, for $j\in\{1,2\}$, put
\[
 R_j=R\cap V(G'_{Y,j}),
 \qquad
 \tau_j=\operatorname{tp}(G'_{Y,j}-R_j).
\]
Each graph $G'_{Y,j}-R_j$ is an induced subgraph of $G'-R$,
so it has treewidth at most $t$.
Moreover, since $Y\subseteq R$,
\[
 (G'_{Y,1}-R_1)\mathbin{\dot\cup}(G'_{Y,2}-R_2)
 =G'-(R\cup Z_Y).
\]
The latter graph is an induced subgraph of $G'-R$.
Consequently,
\[
 \tau_1\oplus\tau_2
 =\operatorname{tp}\bigl(G'-(R\cup Z_Y)\bigr)
 \preceq\operatorname{tp}(G'-R)
 \preceq\tau.
\]
Thus, the algorithm considers the combination of the child
entries indexed by $\tau_1$ and $\tau_2$ as a candidate
for the parent entry indexed by $\tau$.
By the induction hypothesis, these child entries exist and satisfy
\[
 |S_{1,\tau_1}|+|S_{2,\tau_2}|
 \leq
 (1+\delta)(|R_1|+|R_2|)+E_1+E_2,
\]
where $E_j$ is the total separator cost in the comparison
tree witnessing the guarantee for the $j$th child.
The corresponding candidate at the current call is
\[
 D=Y\cup Z_Y\cup S_{1,\tau_1}\cup S_{2,\tau_2}.
\]
As established above, $D$ is feasible for the parent entry.
Its cardinality satisfies
\[
\begin{aligned}
 |D|
 &\leq |Y|+|Z_Y|
       +|S_{1,\tau_1}|+|S_{2,\tau_2}|\\
 &\leq |Y|+(1+\delta)(|R_1|+|R_2|)
       +|Z_Y|+E_1+E_2\\
 &\leq (1+\delta)|R|+|Z_Y|+E_1+E_2.
\end{aligned}
\]
The last inequality follows because $Y,R_1,R_2$ are
pairwise disjoint subsets of $R$.
The comparison tree for the current call consists of its
root and the two child comparison trees, and hence its
total separator cost is
\(
 E=|Z_Y|+E_1+E_2.
\)
Since the algorithm retains a minimum-cardinality candidate
for each entry, its returned entry satisfies
\(
 |S_\tau|\leq |D|\leq (1+\delta)|R|+E.
\)
This completes the induction.

Now apply the guarantee at the root with
$R=S^\star$ and
$\tau^\star=\operatorname{tp}(G-S^\star)$,
where $S^\star$ is an optimum solution.
Using the separator-cost bound and $q\leq a_tk$, we obtain
\[
 |S_{\tau^\star}|
 \leq (1+\delta)k+\frac{Cq}{\log^2 s_0 }
 \leq
 \left(1+\delta+\frac{Ca_t}{\log^2 s_0)}\right)k.
\]
Choose $s_0$ sufficiently large that all earlier requirements
hold and
\(
 \frac{Ca_t}{\log^2 s_0}\leq\frac{\varepsilon}{2}.
\)
Since $\delta=\varepsilon/2$, it follows that
\(
 |S_{\tau^\star}|\leq(1+\varepsilon)k.
\)
Furthermore,
\(
 \operatorname{tw}(G-S_{\tau^\star})\leq t \text{ and }
 \operatorname{tp}(G-S_{\tau^\star})\preceq\tau^\star.
\)
Because $G-S^\star\models\phi$ and $\phi$ is hereditary,
every type below $\tau^\star$ satisfies $\phi$.
Thus, $S_{\tau^\star}$ is a feasible solution to the original
instance. Returning the smallest stored deletion set whose
remainder satisfies $\phi$ therefore yields a feasible
solution of size at most $(1+\varepsilon)k$.

Now we bound the running time. 
The number of states is constant.
At a call with $q'$ marked vertices, Lemma~\ref{lem:qp-local} produces at most
\(
 2^{c_2\delta^{-3}\log^9 q'}
\)
branches, each with two recursive children.
The number of marked vertices decreases by a factor of
$2/3$ along every recursive edge, so the recursion depth is
$O(\log n)$.
Consequently, the total number of recursive calls is at most
\(
 2^{O_{H,t,\varepsilon}(\log^{10} n)}.
\)
All remaining work at each call, including the
bounded-treewidth computations and the constant-size table
combinations, takes polynomial time for fixed
$H,t,\phi,\varepsilon$.
The total running time is therefore
\(
 2^{O_{H,t,\phi,\varepsilon}(\log^{10} n)},
\)
which is quasipolynomial.
\end{proof}

\bibliographystyle{plain}
\bibliography{ref}


\appendix

\section{Proof of Lemma~\ref{lem:degenbranching}} \label{appx:proof}
We mainly follow the proof of~\cite[Lemma~5.1]{korhonen2024induced}
The branching algorithm will maintain a partition $\{A,B,U\}$ of the vertex set $V(G)$, and a ``potential function'' $\phi : V(G) \rightarrow [-1, \delta]$ on vertices.
Each element of the collection $\mathcal{A}$ returned by the algorithm will be the set $A$ in (some of) the leaves of this branching.
Besides, the set $B$ consists of the vertices that we decide to not include in $A$, and the set $U$ consists of the ``undecided'' vertices.
At the beginning, $U = V(G)$ and $\phi(v) = t$ for all vertices $v \in V(G)$.
The intuition will be that the function $\phi(v)$ counts how many more out-neighbors the vertex $v$ is allowed to have in the set $B$.
The algorithm maintains the invariants that $G[B]$ is $t$-degenerate and $\phi(u) \ge 0$ for all $u \in B \cup U$; whenever one of them is violated, the current branch is terminated without output.

Next we describe the branching.
First, if every vertex of $U$ already has degree at most $\Delta$ in $G[B \cup U]$, then we add $A$ to the collection $\mathcal{A}$ (and terminate the branch).
Otherwise, there exists $v \in U$ so that $v$ has degree more than $\Delta$ in $G[B \cup U]$.
We arbitrarily select such a vertex $v$, and branch on it in the following sense.
In the first branch, we simply move $v$ from $U$ to $A$.
In the second branch, we move $v$ from $U$ to $B$ and further branch on all possibilities of a subset $N_\mathsf{out}(v) \subseteq N_G(v) \cap (B \cup U)$ with $|N_\mathsf{out}| \leq t$, and in each of these branches, decrease $\phi(u)$ by one for all vertices $u \in (N_G(v) \cap (B \cup U)) \backslash N_\mathsf{out}(v)$.
After that, if $G[B]$ is not $t$-degenerate or $\phi(u) = -1$ for some $u \in B$, we terminate the branch without output.
Otherwise, we move all vertices $u \in U$ with $\phi(u) = -1$ to the set $A$.
Note that the $\phi$-value of each vertex is decreased at most once per step, so $\phi(u) \ge -1$ always holds.

To see the correctness of the above branching algorithm, we first show the collection $\mathcal{A}$ returned satisfies condition (i).
Let $A \in \mathcal{A}$ and consider the leaf of the branching tree at which $A$ is added to $\mathcal{A}$.
By the invariant, at this point, $G[B]$ is $t$-degenerate, and every vertex of $U = V(G) \backslash (A \cup B)$ has degree at most $\Delta$ in $G[B \cup U] = G - A$.
It remains to bound $|B|$.
To this end, we consider the function $\phi^+ : V(G) \rightarrow [0,\delta+1]$ defined as $\phi^+(v) = \phi(v)+1$ for all $v \in V(G)$, and let $\phi^+(B \cup U) = \sum_{v \in B \cup U} \phi^+(v)$.
Observe that at the beginning, the value of $\phi^+(I \cup U)$ is $(t+1) \cdot n$, that it is always non-negative, and that moving vertices from $U$ to $A$ does not increase it.
In all branches of the second type, the vertex $v$ stays in $B \cup U$, and $\phi^+(u)$ decreases by one for all $u \in (N_G(v) \cap (B \cup U)) \backslash N_\mathsf{out}(v)$, so the value $\phi^+(B \cup U)$ decreases by at least $\Delta+1-t$.
Therefore, in any root-leaf path of the branching tree, there can be at most $(t+1)n/(\Delta-t+1)$ branches of the second type.
This implies that in each leaf, the set $B$ is of size at most $(t+1)n/(\Delta-t+1)$.

Then we show that $\mathcal{A}$ satisfies condition (ii).
Consider a subset $S \subseteq V(G)$ such that $G-S$ is $t$-degenerate.
We need to show the existence of $A \in \mathcal{A}$ with $A \subseteq S$.
Let $\sigma$ be a fixed $t$-degeneracy ordering of $G-S$.
For $v \in V(G) \backslash S$, denote by $\varGamma(v)$ the set of (at most $t$) out-neighbors of $v$ in $G-S$ under $\sigma$.
We say a node of the branching tree is \textit{good} if at the node we have $A \subseteq S$, $B \cap S = \emptyset$, and $\phi(v) \geq |\varGamma(v) \cap U|$ for all $v \in V(G) \backslash S$.
Note that the root of the branching tree is good.
Furthermore, we define the notion of \textit{correct} and \textit{wrong} branches as follows.
During the algorithm, when branching on a vertex $v \in U$,
\begin{itemize}
    \item if $v \in S$, then the branch that moves $v$ to $A$ is correct (and all the other branches are wrong),
    \item if $v \notin S$, then the branch that moves $v$ to $B$ and chooses the subset $N_\mathsf{out}(v) = \varGamma(v) \backslash S$ is correct  (and all the other branches are wrong).
\end{itemize}
Observe that at a good node of the branching tree, exactly one of its branches is correct, because $|\varGamma(v) \backslash S| \leq t$ and for a good node we have $\varGamma(v) \backslash S \subseteq N_G(v) \backslash A = N_G(v) \cap (B \cup U)$.
Furthermore, one can easily verify that if $y$ is a child of a good node $x$ and the branch $x \rightarrow y$ is correct, then $y$ is also good.
Therefore, if we go from the root of the branching tree and follow the correct branches, we will reach a leaf that is good.
At this leaf, the set $A$ is added to $\mathcal{A}$ and it holds that $A \subseteq S$.

Finally, it suffices to bound the time complexity and the size of $\mathcal{A}$.
At each node of the branching tree, at most one branch is of the first type and $n^{O(t)}$ branches are of the second type.
Furthermore, as observed before, along any root-leaf path, there can be at most $(t+1)n/(\Delta-t+1) = O(tn/\Delta)$ branches of the second type.
Since the height of the branching tree is at most $n$,  the number of root-leaf paths is at most 
\begin{align*}
n^{O(tn/\Delta)} \cdot n^{O(t^2 n/\Delta)} = n^{O(t^2 n/\Delta)}.
\end{align*}
This directly bounds the number of nodes in the branching tree.
Since we add at most one set to $\mathcal{A}$ at each leaf and the time cost at each node is $n^{O(1)}$, we see that $|\mathcal{A}| = n^{O(t^2 n/\Delta)}$ and the time complexity of the algorithm is also $n^{O(t^2 n/\Delta)}$.

\end{document}

%% file: preamble.tex
\documentclass[letter,11pt]{article}

\usepackage{color}
\usepackage{xcolor}
\usepackage{xspace}
\usepackage{setspace}
\usepackage{subcaption}
\usepackage{enumerate}
\usepackage{amsmath}
\usepackage{amsfonts}
\usepackage{graphicx}
\usepackage{amsthm}
\usepackage{thm-restate}
\usepackage{tcolorbox}

\usepackage{xspace}
\usepackage{amssymb}
\usepackage{multirow}
\usepackage{geometry}
\usepackage{soul}
\usepackage{colortbl}
\usepackage{wrapfig}

\usepackage{algorithm}
\usepackage{algorithmicx}
\usepackage{algpseudocode}

\usepackage{todonotes}
\usepackage{mathtools}
\usepackage{mathrsfs}
\usepackage{subcaption}

\usepackage[pdftex, plainpages = false, pdfpagelabels, 
                 bookmarks=false,
                 bookmarksopen = true,
                 bookmarksnumbered = true,
                 breaklinks = true,
                 linktocpage,
                 pagebackref,
                 colorlinks = true,  
                 linkcolor = blue,
                 urlcolor  = blue,
                 citecolor = red,
                 anchorcolor = green,
                 hyperindex = true,
                 hyperfigures
                 ]{hyperref}

\usepackage{aliascnt}

\newtheorem{lemma}{Lemma}

\numberwithin{lemma}{section}

\newtheorem{theorem*}[lemma]{Theorem*}
\newtheorem{corollary}[lemma]{Corollary}

\newtheorem{observation}[lemma]{Observation}

\newcommand*{\claimproofs}{Proof of the Claim.}
\newenvironment{claimproof}[1][\claimproofs]{\begin{proof}[#1]}{\end{proof}}

\usetikzlibrary{arrows.meta,positioning,calc,fit}

\AtBeginDocument{%
  \let\originaltableofcontents\tableofcontents
  \renewcommand{\tableofcontents}{\par\noindent
    \begin{minipage}{\linewidth}\originaltableofcontents\end{minipage}\par}}

%% file: authors.tex
    \title{Vertex Deletion on $H$-Induced-Minor-Free Graphs:\protect\\ Subexponential FPT Algorithms and Approximation Schemes}

\author{%
Daniel Lokshtanov\thanks{%
University of California, Santa Barbara, USA.
Email: \texttt{daniello@ucsb.edu}.}
\and
Fahad Panolan\thanks{%
University of Leeds, UK.
Email: \texttt{F.Panolan@leeds.ac.uk}.}
\and
Saket Saurabh\thanks{%
The Institute of Mathematical Sciences, HBNI, Chennai, India;
and University of Bergen, Norway.
Email: \texttt{saket@imsc.res.in}.}
\and
Jie Xue\thanks{%
New York University Shanghai, China.
Email: \texttt{jiexue@nyu.edu}.}
\and
Meirav Zehavi\thanks{%
Ben-Gurion University of the Negev, Israel.
Email: \texttt{meiravze@bgu.ac.il}.}
}
\date{}

%% file: frontmatter.tex
\begin{abstract}
We study vertex-deletion problems on string graphs and, more generally, on graphs excluding a fixed graph as an induced minor. For every fixed graph $H$, fixed nonnegative integer $t$, and fixed monadic second-order formula $\phi$, we consider the problem of finding a minimum-cardinality set $S$ such that $G-S$ has treewidth at most $t$ and satisfies $\phi$. This captures classic problems such as \textsc{Vertex Cover}, \textsc{Feedback Vertex Set}, and deletion to graphs of any fixed bounded treewidth.
We give an algorithm with running time $2^{k^{2/3} \log^{O(1)} k}n^{O(1)}$ on $H$-induced-minor-free graphs, where $k$ is the optimum deletion-set size. 
This resolves in the affirmative the open problem highlighted by Bonnet and {\fontencoding{T1}\selectfont Rz\k{a}\.{z}ewski} [Algorithmica~2019] and Berthe, Bougeret, Gon\c{c}alves, and Raymond [ICALP~2025] regarding the existence of a subexponential parameterized algorithm for \textsc{Feedback Vertex Set} on string graphs. 
Our algorithm significantly improves over existing parameterized algorithms for {\em more restricted} classes of problems on {\em more restricted} classes of graphs.

When the property defined by $\phi$ is hereditary, we also obtain a quasi-polynomial time approximation scheme, namely an algorithm that produces a $(1+\varepsilon)$-approximate solution in quasi-polynomial time.
Both the sub-exponential algorithm and the quasi-polynomial time approximation scheme build on the same new structural and algorithmic insights. 

A key ingredient that may be of independent interest is a new covering-packing theorem that works for general graphs:
If $G$ has a vertex set $S$ such that $G-S$ has treewidth at most $t$, then for every integer $r$ we can either find $2k$ pairwise disjoint and anti-complete connected subsets, each with at least $2r$ neighbors in $S$, or we can find a set $S'$ containing $S$ of size at most $O(rtk)$ such that every component of $G-S'$ has at most $O(r)$ neighbors in $S'$. 




\end{abstract}


%% file: algo.tex
\subsection{The algorithm} \label{sec:algo}

In this section, we present the branching algorithm of Theorem~\ref{thm:branching}.
For convenience of analysis, instead of computing a collection $\mathcal{X} = \{X_1,\dots,X_r\}$ of subsets of $V(G)$ as in Theorem~\ref{thm:branching}, our actual algorithm will compute a collection $\mathcal{Q} = \{(X_1,M_1),\dots,(X_r,M_r)\}$ of pairs of subsets of $V(G)$.
Specifically, the main goal of this section is to prove the following lemma.

\begin{lemma} \label{lem:branching}
Given a graph $G$ with $n$ vertices and three parameters $k,d,t \in \mathbb{N}$ with $d > 2t$, one can compute in $k^{t^{O(1)} k/d} n^{O(1)}$ time a collection $\mathcal{Q} = \{(X_1,M_1),\dots,(X_r,M_r)\}$ where $r = k^{t^{O(1)} k/d}$ and $X_1,\dots,X_r,M_1,\dots,M_r \subseteq V(G)$ satisfy the following conditions:
\begin{enumerate}[(i)]
    \item $\mathbf{tw}(G-(X_i \cup M_i)) \leq t$ and $|M_i| = O(t^2 \log t \cdot k)$ for all $i \in [r]$,
    \item $M_i$ is an $O(d+t)$-core of $G- X_i$ for all $i \in [r]$,
    \item $G[M_i]$ is $d$-degenerate for $i \in [r]$,
    \item for every treewidth-$t$ modulator $S \subseteq V(G)$ of $G$ satisfying $|S| \leq k$, there exists an index $i \in [r]$ such that $X_i \subseteq S$.
\end{enumerate}
\end{lemma}

The algorithm of Theorem~\ref{thm:branching} is the same as the one in the above lemma, which returns $\mathcal{X} = \{X_1,\dots,X_r\}$ extracted from the collection $\mathcal{Q} = \{(X_1,M_1),\dots,(X_r,M_r)\}$ in the lemma.
Note that (iv) of Lemma~\ref{lem:branching} directly implies (ii) of Theorem~\ref{thm:branching}.
However, the treewidth bound in (i) of Theorem~\ref{thm:branching} only holds when $G \in \mathcal{G}_H$ for some fixed $H$, and proving this is way more difficult.
To this end, we need to establish several new structural results for induced-minor-free graphs, and apply them together with (i), (ii), (iii) of Lemma~\ref{lem:branching}.

To prove Lemma~\ref{lem:branching}, we first discuss the sub-routines used in our algorithm.
The first one we need is the approximation algorithm for minimum treewidth modulator in Lemma~\ref{lem:twmod}; we denote this algorithm by $\textsc{TwMod}(G,t)$, whose output is a treewidth-$t$ modulator of the graph $G$.
The second one we need is the algorithm in Corollary~\ref{cor:decomp}; we denote this algorithm by $\textsc{Decompose}(G,M,d,t)$, whose output is the collection $\mathcal{P}$in the corollary.
Furthermore, we need the algorithm in Lemma~\ref{lem:closure}; we denote this algorithm by $\textsc{Closure}(G,Z)$, whose output is the set $L \subseteq V(G)$ in the lemma.
Finally, we also need the algorithm in the following lemma, which is a strengthening of a lemma by Korhonen and Lokshtanov~\cite{korhonen2024induced} (see Lemma~5.1 in the arxiv version).

\begin{lemma}[\cite{korhonen2024induced}] \label{lem:degenbranching}
Given a graph $G$ with $n$ vertices and parameters $\Delta,t \in \mathbb{N}$ with $\Delta \geq t$, one can compute in $n^{O(t^2 \cdot n/\Delta)}$ time a collection $\mathcal{A} = \{A_1,\dots,A_r\}$ where $r = n^{O(t^2 \cdot n/\Delta)}$ and $A_1,\dots,A_r \subseteq V(G)$ satisfy the following conditions:
\begin{enumerate}[(i)]
    \item for each $i \in [r]$, there exists $B_i \subseteq V(G) \backslash A_i$ such that
    \begin{itemize}
        \item $G[B_i]$ is $t$-degenerate,
        \item the degree of every vertex $v \in V(G) \backslash (A_i \cup B_i)$ in $G - A_i$ is at most $\Delta$,
    \end{itemize}
    in particular, $G-A_i$ is $\Delta$-degenerate,
    \item for every subset $S \subseteq V(G)$ satisfying that and $G-S$ is $t$-degenerate, there exists an index $i \in [r]$ such that $A_i \subseteq S$.
\end{enumerate}
\end{lemma}

The lemma can be proved in very much the same way as~\cite[Lemma~5.1]{korhonen2024induced}; for completeness, we give a proof in Appendix~\ref{appx:proof}.
We denote the algorithm in the lemma by $\textsc{Degen}(G,\Delta,t)$, whose output is the collection $\mathcal{A}$ in the lemma.

Now we are ready to discuss the branching algorithm of Lemma~\ref{lem:branching}, which is presented in Algorithm~\ref{alg-branch1}.
First, we apply Lemma~\ref{lem:twmod} to compute a treewidth-$t$ modulator $M_\mathsf{init}$ of the graph $G$.
The desired collection $\mathcal{Q}$ is computed via a recursive procedure $\textsc{Branch}$, which is the technical core of our algorithm.
Each call of $\textsc{Branch}$ has two arguments, $X$ and $M$, which are disjoint subsets of $V(G)$.
We set $\mathcal{Q} = \textsc{Branch}(X = \emptyset,M=M_\mathsf{init})$.

Next, we explain how $\textsc{Branch}(X,M)$ works.
If $|X| > k$, then it simply returns the empty collection $\emptyset$.
Otherwise, we apply Corollary~\ref{cor:decomp} to compute $\mathcal{P} = \textsc{Decompose}(G-X,M,d,t)$, which satisfies the four conditions in the corollary.
Each element in $\mathcal{P}$ is a pair $(V,Z)$, where $V,Z \subseteq V(G) \backslash M$.
The algorithm distinguishes two cases: $|\mathcal{P}| \leq |M|+k$ and $|\mathcal{P}| > |M|+k$.
The former case corresponds to line~7-11 and the latter case corresponds to line~12-17 in Algorithm~\ref{alg-branch1}.

If $|\mathcal{P}| \leq |M|+k$, the procedure $\textsc{Branch}$ will directly construct a collection $\mathcal{Q}$ and return it, as follows.
First, we apply Lemma~\ref{lem:closure} to compute $L = \textsc{Closure}(G-(X \cup M),\bigcup_{(V,Z) \in \mathcal{P}} Z)$, which is a subset of $V(G) \backslash (X \cup M)$ that contains $\bigcup_{(V,Z) \in \mathcal{P}} Z$.
Then Lemma~\ref{lem:degenbranching} is applied to compute $\mathcal{A} = \textsc{Degen}(G[M \cup L],d,t)$, which is a collection of subsets of $M \cup L$.
The collection $\mathcal{Q}$ to be returned consists of pairs $(X \cup A, (M \cup L) \backslash A)$ for all $A \in \mathcal{A}$.

On the other hand, if $|\mathcal{P}| > |M|+k$, we need to recursively call $\textsc{Branch}(X \cup X',M \backslash X')$ for several subsets $X' \subseteq M$.
Specifically, we take a sub-collection $\mathcal{P}' \subseteq \mathcal{P}$ of size $|M|+k$.
We consider all pairs $(V,Z) \in \mathcal{P}'$.
For each $(V,Z) \in \mathcal{P}'$, we call $\textsc{Branch}(X \cup X',M \backslash X')$ for all subsets $X' \subseteq N_G(V) \cap M$ satisfying $|(N_G(V) \cap M) \backslash X'| \leq 2t$.
The collection $\mathcal{Q}$ to be returned is the union of the collections returned by these recursive calls.

\begin{algorithm}[h]
    \caption{\textsc{Core}$(G,k,d,t)$}
    \begin{algorithmic}[1]
        \State $M_\mathsf{init} \leftarrow \textsc{TwMod}(G,t)$ \Comment{Lemma~\ref{lem:twmod}}
        \State $\mathcal{Q} \leftarrow \textsc{Branch}(\emptyset,M_\mathsf{init})$
        \State \textbf{return} $\mathcal{Q}$
        \smallskip
        \Function{\textsc{Branch}}{$X,M$}
            \If{$|X| > k$}{ \textbf{return} $\emptyset$}
            \EndIf
            \State $\mathcal{P} \leftarrow \textsc{Decompose}(G-X,M,d,t)$ \Comment{Corollary~\ref{cor:decomp}}
            \If{$|\mathcal{P}| \leq |M|+k$}
                \State $L \leftarrow \textsc{Closure}(G-(X \cup M),\bigcup_{(V,Z) \in \mathcal{P}} Z)$ \Comment{Lemma~\ref{lem:closure}}
                \State $\mathcal{A} \leftarrow \textsc{Degen}(G[M \cup L],d,t)$ \Comment{Lemma~\ref{lem:degenbranching}}
                \State $\mathcal{Q} \leftarrow \{(X \cup A,(M \cup L) \backslash A): A \in \mathcal{A}\}$
                \State \textbf{return} $\mathcal{Q}$
            \Else
            \State $\mathcal{P}' \leftarrow$ a sub-collection of $\mathcal{P}$ with size $|M|+k$
            \State $\mathcal{Q} \leftarrow \emptyset$
            \For{every $(V,Z) \in \mathcal{P}'$}
                \For{every $X' \subseteq N_G(V) \cap M$ satisfying $|(N_G(V) \cap M) \backslash X'| \leq 2t$}
                    \State $\mathcal{Q} \leftarrow \mathcal{Q} \cup \textsc{Branch}(X \cup X',M \backslash X')$
                \EndFor
            \EndFor
            \State \textbf{return} $\mathcal{Q}$
            \EndIf
        \EndFunction
    \end{algorithmic}
    \label{alg-branch1}
\end{algorithm}

To analyze Algorithm~\ref{alg-branch1}, let $T_\mathsf{rec}$ be the recurrence tree of the call $\textsc{Branch}(\emptyset,M_\mathsf{init})$ in line~2 of Algorithm~\ref{alg-branch1}.
The root of $T_\mathsf{rec}$ corresponds to the initial call, and each non-root node of $T_\mathsf{rec}$ corresponds to a recursive call of $\textsc{Branch}$.
For a node $x \in V(T_\mathsf{rec})$, we denote by $X_x$ and $M_x$ the arguments $X$ and $M$ of the call corresponding to $x$, respectively.
Thus, $X_\mathsf{rt} = \emptyset$ and $M_\mathsf{rt} = M_\mathsf{init}$ for the root $\mathsf{rt}$ of $T_\mathsf{rec}$.
Also, let $\mathcal{Q}_x$ denote the collection returned by $\textsc{Branch}(X_x,M_x)$.
Clearly, if $x$ is a non-leaf node, then $\mathcal{Q}_x = \bigcup_{y \in \mathsf{ch}(x)} \mathcal{Q}_y$, where $\mathsf{ch}(x)$ is the set of the children of $x$.
As such, the final collection $\mathcal{Q}$ returned by Algorithm~\ref{alg-branch1} is just $\mathcal{Q} = \bigcup_{x \in V(T_\mathsf{rec})} \mathcal{Q}_x$.

\begin{observation} \label{obs:disjoint}
    $X_x \cap M_x = \emptyset$ and $X_x \cup M_x = M_\mathsf{init}$ for all $x \in V(T_\mathsf{rec})$.
\end{observation}
\begin{proof}
Consider a call $\textsc{Branch}(X,M)$.
All its recursive calls are of the form $\textsc{Branch}(X \cup X',M \backslash X')$, where $X' \subseteq M$.
Thus, if $X_x \cap M_x = \emptyset$ and $X_x \cup M_x = M_\mathsf{init}$ for a node $x \in V(T_\mathsf{rec})$, then $X_y \cap M_y = \emptyset$ and $X_y \cup M_y = M_\mathsf{init}$ for all children of $x$ in $T_\mathsf{rec}$.
The root $\mathsf{rt}$ of $T_\mathsf{rec}$ clearly satisfies $X_\mathsf{rt} \cap M_\mathsf{rt} = \emptyset$ and $X_\mathsf{rt} \cup M_\mathsf{rt} = M_\mathsf{init}$.
As such, we have the property for all $x \in V(T_\mathsf{rec})$.
\end{proof}

\begin{observation} \label{obs:depth}
    Let $x$ be a node at the $\ell$-th level of $T_{\mathsf{rec}}$.
    Then $|X_x| \geq \ell (d-2t)$.
\end{observation}
\begin{proof}
We apply induction on $\ell$.
When $\ell = 0$, the statement trivially holds.
Assume that for all nodes $x$ at the $(\ell-1)$-th level of $T_{\mathsf{rec}}$, $|X_x| \geq (\ell-1)(d-2t)$.
To prove the statement for the nodes at the $\ell$-th level, consider a node $x$ at the $(\ell-1)$-th level and we need to show $|X_y| \geq \ell (d-2t)$ for all children $y$ of $x$.
Note that for each child $y$ of $x$, we have $X_y = X_x \cup X'$ for some $X'$ chosen in line~16 of the call the call $\textsc{Branch}(X_x,M_x)$.
So it suffices to show $|X'| \geq d-2t$ for all $X'$ chosen in line~16.
By (ii) of Corollary~\ref{cor:decomp}, $|N_G(V) \cap M| > d$ for all $(V,Z) \in \mathcal{P}$ where $\mathcal{P}$ is the collection computed in line~6.
Since $X'$ is a subset of $N_G(V) \cap M$ for some $(V,Z) \in \mathcal{P}' \subseteq \mathcal{P}$ and $|(N_G(V) \cap M) \backslash X'| \leq 2t$ as required in line~16, we have $|X'| \geq d-2t$.
Therefore, $|X_y| \geq \ell (d-2t)$ for all children $y$ of $x$.
\end{proof}

To prove Lemma~\ref{lem:branching}, we first show that the collection $\mathcal{Q}$ satisfy the four desired conditions in the lemma, and then bound the time complexity of the algorithm as well as the size of $\mathcal{Q}$.

\paragraph{Correctness.}
To see $\mathcal{Q}$ satisfies the desired conditions, we observe that every element in $\mathcal{Q}$ is generated via line~10 of Algorithm~\ref{alg-branch1} in some call corresponding to a leaf of $T_{\mathsf{rec}}$.

To verify condition (i), we want to show $\mathbf{tw}(G - (X \cup A \cup M \cup L)) \leq t$ and $|(M \cup L) \backslash A| = O(t^2 \log t \cdot k)$ for all $A \in \mathcal{A}$ whenever the algorithm executes line~10.
It suffices to show $\mathbf{tw}(G - (X \cup M)) \leq t$ and $|M \cup L| = O(t^2 \log t \cdot k)$.
We have $X \cup M = M_\mathsf{init}$ by Observation~\ref{obs:disjoint}, which implies $\mathbf{tw}(G - (X \cup M)) \leq t$ since $M_\mathsf{init}$ is a treewidth-$t$ modulator of $G$.
To bound the size of $M \cup L$, note that $|M| \leq |M_\mathsf{init}| = O(k \log t)$ by Observation~\ref{obs:disjoint} and Lemma~\ref{lem:twmod}.
By (iii) of Corollary~\ref{cor:decomp}, the collection $\mathcal{P}$ constructed in line~6 satisfies $|Z| = O(t)$ for all $(V,Z) \in \mathcal{P}$.
Since line~10 is only executed when $|\mathcal{P}| \leq |M| + k$, we further have $|\bigcup_{(V,Z) \in \mathcal{P}} Z| = O(t(|M|+k))=O(t\log t \cdot k)$.
Then by Lemma~\ref{lem:closure}, $|L| = O(t|\bigcup_{(V,Z) \in \mathcal{P}} Z|) = O(t^2 \log t \cdot k)$, because $\mathbf{tw}(G - (X \cup M)) \leq t$.
Therefore, $|M \cup L| = O(t^2 \log t \cdot k)$.
Condition (i) holds.

To verify condition (ii), we need to show that $(M \cup L) \backslash A$ is an $O(d+t)$-core of $G - (X \cup A)$ for all $A \in \mathcal{A}$ when executing line~10.
Again, consider the collection $\mathcal{P}$ constructed in line~6.
By (iv) of Corollary~\ref{cor:decomp}, $M$ is a $d$-core of $G - (X \cup (\bigcup_{(V,Z) \in \mathcal{P}} Z))$, and thus a $d$-core of $G - (X \cup L)$.
Exploiting the fact $\mathbf{tw}(G - (X \cup M)) \leq t$, we see $L$ is an $O(t)$-core of $G - (X \cup M)$ by Lemma~\ref{lem:closure}.
Therefore, $M \cup L$ is an $O(d+t)$-core of $G - X$ by Lemma~\ref{lem:core2}.
Now Lemma~\ref{lem:core1} further implies that $(M \cup L) \backslash A$ is an $O(d+t)$-core of $G - (X \cup A)$ for all $A \in \mathcal{A}$.
So condition (ii) holds.

To verify condition (iii), we need to show that $G[(M \cup L) \backslash A]$ is $d$-degenerate for all $A \in \mathcal{A}$ when executing line~10.
This directly follows from (i) of Lemma~\ref{lem:degenbranching}, since $\mathcal{A} = \textsc{Degen}(G[M \cup L],d,t)$, i.e., we apply the algorithm of Lemma~\ref{lem:degenbranching} on the graph $G[M \cup L]$ with $\Delta = d$. 

Finally, we verify condition (iv).
Let $S \subseteq V(G)$ be a treewidth-$t$ modulator of $G$ with $|S| \leq k$.
We need to show there exists $(X,M) \in \mathcal{Q}$ such that $X \subseteq S$.
We say a node $x \in V(T_{\mathsf{rec}})$ is \textit{$S$-good} if $X_x \subseteq S$.
The following observations for $S$-good nodes is crucial.
\begin{observation} \label{obs:Sgood}
    For an $S$-good node $x \in V(T_{\mathsf{rec}})$, either the call $\textsc{Branch}(X_x,M_x)$ returns in line~11 or $x$ has a child in $T_{\mathsf{rec}}$ that is also $S$-good.
\end{observation}
\begin{proof}
Since $x$ is $S$-good, we have $|X_x| \leq |S| \leq k$.
Thus, $\textsc{Branch}(X_x,M_x)$ cannot return in line~5.
Assume it does not return in line~11 either.
We want to show that $x$ has an $S$-good child.
Consider the collection $\mathcal{P}$ computed in line~6 in the call $\textsc{Branch}(X_x,M_x)$.
Since the call does not return in line~11, we have $|\mathcal{P}| > |M|+k$.
In line~13, we construct the subset $\mathcal{P}' \subseteq \mathcal{P}$ of size $|M|+k$.
Write $\mathcal{P}' = \{(V_1,Z_1),\dots,(V_{|M|+k},Z_{|M|+k})\}$.
By (i) of Corollary~\ref{cor:decomp}, $V_1,\dots,V_{|M|+k}$ are disjoint.
Since $|S| \leq k$, $S$ intersects at most $k$ sets among $V_1,\dots,V_{|M|+k}$.
Without loss of generality, we can assume $V_i \cap S = \emptyset$ for all $i \in [|M|]$.
By (ii) of Corollary~\ref{cor:decomp}, $G[V_i]$ is connected for all $i \in [|M|]$.
Let $H$ be the graph obtained from $G$ by contracting each $V_i$ into a single vertex $v_i$ for $i \in [|M|]$.
Note that $H-S$ is a minor of $G-S$, and thus $\mathbf{tw}(H-S) \leq t$.
Define $H' = H[(M \backslash S) \cup \{v_1,\dots,v_{|M|}\}]$.
We have $|V(H')| \leq 2|M|$ and $\mathbf{tw}(H') \leq t$.
Therefore, $|E(H')| \leq 2t |M|$.
Observe that $\sum_{i=1}^{|M|} |(N_{H'}(v_i) \cap M) \backslash S| = \sum_{i=1}^{|M|} |N_{H'}(v_i) \cap (M \backslash S)| \leq |E(H')|$.
As such, there exists $i \in [|M|]$ such that $|(N_{H'}(v_i) \cap M) \backslash S| \leq |E(H')|/|M| \leq 2t$.
Note that $N_{H'}(v_i) \cap M = N_G(V_i) \cap M$.
Let $X' = N_G(V_i) \cap M \cap S$.
We then have
\begin{equation*}
    |(N_G(V_i) \cap M) \backslash X'| = |(N_G(V_i) \cap M) \backslash S| = |(N_{H'}(v_i) \cap M) \backslash S| \leq 2t.
\end{equation*}
Hence, when we take $(V_i,Z_i) \in \mathcal{P}'$ in line~15, the set $X'$ is considered in line~16, which implies that there exists a child $y$ of $x$ in $T_{\mathsf{rec}}$ with $X_y = X_x \cup X'$.
We have $X_x \subseteq S$ and $X' \subseteq S$ by construction.
It follows that $X_y \subseteq S$ and $y$ is $S$-good.
\end{proof}

\begin{observation} \label{obs:leaf}
    If a leaf $x$ of $T_{\mathsf{rec}}$ is $S$-good, then there exists $(X,M) \in \mathcal{Q}_x$ such that $X \subseteq S$.
\end{observation}
\begin{proof}
Consider the call $\textsc{Branch}(X_x,M_x)$.
As a leaf, $x$ does not have any children in $T_{\mathsf{rec}}$.
Observation~\ref{obs:Sgood} then implies that $\textsc{Branch}(X_x,M_x)$ returns in line~11.
In line~8, we apply Lemma~\ref{lem:closure} on $G - (X_x \cup M_x)$, and the set $L$ returned is a subset of $V(G) \backslash (X_x \cup M_x)$.
Then in line~9, we apply Lemma~\ref{lem:degenbranching} on the graph $G[M_x \cup L]$, and the collection $\mathcal{A}$ returned consists of subsets of $M_x \cup L$.
As $S$ is a treewidth-$t$ modulator of $G$, $\mathbf{tw}(G-S) \leq t$ and thus $G-S$ is $t$-degenerate.
It follows that $G[M_x \cup L] - (S \cap (M_x \cup L))$ is also $t$-degenerate.
Therefore, (ii) of Lemma~\ref{lem:degenbranching} implies there exists $A \in \mathcal{A}$ such that $A \subseteq S \cap (M_x \cup L) \subseteq S$.
Since $x$ is $S$-good, we have $X_x \subseteq S$ and hence $X_x \cup A \subseteq S$.
Note that $(X_x \cup A,(M_x \cup L) \backslash A) \in \mathcal{Q}_x$ by line~10, which completes the proof.
\end{proof}

To see $\mathcal{Q}$ consists of a pair $(X,M)$ satisfying $X \subseteq S$, observe that $T_{\mathsf{rec}}$ contains at least one node that is $S$-good, simply because the root of $T_{\mathsf{rec}}$ is $S$-good.
Let $x$ be a lowest $S$-good node in $T_{\mathsf{rec}}$.
By definition, $x$ does not have any $S$-good child.
Observation~\ref{obs:Sgood} implies that the call $\textsc{Branch}(X_x,M_x)$ returns in line~9 and thus $x$ is a leaf of $T_\mathsf{rec}$.
Then by Observation~\ref{obs:leaf}, there exists $(X,M) \in \mathcal{Q}_x$ such that $X \subseteq S$.
Since $\mathcal{Q}_x \subseteq \mathcal{Q}$, condition (iv) holds.

\paragraph{Time complexity.}
Next, we analyze the time complexity of Algorithm~\ref{alg-branch1} and the size of the returned collection $\mathcal{Q}$.
Line~1 takes $2^{t^{O(1)}} n^{O(1)}$ time by Lemma~\ref{lem:twmod}.
Applying the algorithms of Corollary~\ref{cor:decomp} and Lemma~\ref{lem:closure} in each call of \textsc{Branch} takes $2^{O(t)} n^{O(1)}$ time.
The size of the argument $M$ is bounded by $|M_\mathsf{init}| = O(k \log t)$.
Furthermore, as observed before, the set $L$ computed in line~8 is of size $O(t\log t \cdot k)$.
Therefore, applying the algorithm of Lemma~\ref{lem:degenbranching} in line~9 takes $(t \log t \cdot k)^{O(t^3 \log t \cdot k/d)} n^{O(1)}$ time, and the returned collection $\mathcal{A}$ is of size $(t \log t \cdot k)^{O(t^3 \log t \cdot k/d)}$.
So the time cost of the call $\textsc{Branch}(\emptyset,M_\mathsf{init})$ in line~2 can be bounded by $k^{t^{O(1)} k/d} n^{O(1)} \cdot |V(T_{\mathsf{rec}})|$.
In addition, the size of $\mathcal{Q}_x$ for each leaf $x$ of $T_{\mathsf{rec}}$ is bounded by $k^{t^{O(1)} k/d}$, which then implies $|\mathcal{Q}| = k^{t^{O(1)} k/d} \cdot |V(T_{\mathsf{rec}})|$.
It now suffices to bound $|V(T_{\mathsf{rec}})|$.

We first notice that the degree of $T_{\mathsf{rec}}$ is $(k \log t)^{O(t)}$.
The collection $\mathcal{P}'$ computed in line~13 is of size $|M|+k = O(k \log t)$.
For each $(V,Z) \in \mathcal{P}'$, the number of $X' \subseteq N_G(V) \cap M$ satisfying $|(N_G(V) \cap M) \backslash X'| \leq 2t$ is bounded by $|N_G(V) \cap M|^{O(t)} \leq |M|^{O(t)} = (k \log t)^{O(t)}$.
As such, line~17 is executed $(k \log t)^{O(t)}$ time in each call of $\textsc{Branch}$, and thus the degree of $T_{\mathsf{rec}}$ is $(k \log t)^{O(t)}$.
Next, we show the depth of $T_{\mathsf{rec}}$ is at most $k/(d-2t)+1$.
If a node $x \in V(T_{\mathsf{rec}})$ satisfies $|X_x| > k$, then the call $\textsc{Branch}(X_x,M_x)$ returns in line~5 and hence $x$ must be a leaf of $T_{\mathsf{rec}}$.
So we have $|X_x| \leq k$ for all internal nodes $x \in V(T_{\mathsf{rec}})$.
By Observation~\ref{obs:depth}, this implies the depth of an internal node in $T_{\mathsf{rec}}$ is at most $k/(d-2t)$.
Therefore, the depth of $T_{\mathsf{rec}}$ is at most $k/(d-2t)+1 = O(t k/d)$.
It follows that $|V(T_{\mathsf{rec}})| = k^{t^{O(1)} k/d}$, which completes the proof of Lemma~\ref{lem:branching}.

%% file: torso.tex

We first prove the following structural lemma, which states that when applying the \textsc{Torso} operation to the complement of an independent set in an induced-minor-free graph, the resulting graph preserves the induced-minor-freeness.

\begin{lemma} \label{lem:indeptorso}
For every graph $H$, there exists a graph $H'$ satisfying the following:
\begin{itemize}
    \item For any $G \in \mathcal{G}_H$ and any independent set $I \subseteq V(G)$ in $G$, we have $\textsc{Torso}(G,V(G) \backslash I) \in \mathcal{G}_{H'}$.
\end{itemize}
\end{lemma}
\begin{proof}
Thanks to Lemma~\ref{lem:subdivclique}, it suffices to prove the lemma for the case $H = K_h^\bullet$ for some number $h \in \mathbb{N}$.
Let $G \in \mathcal{G}_H$ where $H = K_h^\bullet$ and $I \subseteq V(G)$ be an independent set in $G$.
Define $G' = \textsc{Torso}(G,V(G) \backslash I)$.
We shall show that $G' \in \mathcal{G}_H$.
To this end, we define
\begin{equation*}
    \textsc{Ext}(V) = V \cup \{v \in I: N_G[v] \cap V \neq \emptyset\},
\end{equation*}
for $V \subseteq V(G)$.
We say two sets $U,V \subseteq V(G)$ \textit{touch} in $G$ (resp., $G'$) if $U \cap V \neq \emptyset$ or $U$ and $V$ are adjacent in $G$ (resp., $G'$).
We observe the following good properties of \textsc{Ext}.

\medskip

\noindent
\textbf{Claim 1.} \textit{For any $V \subseteq V(G)$, if $G'[V]$ is connected, then $G[\textsc{Ext}(V)]$ is connected as well.}
\begin{claimproof}
It suffices to show that for all $u,v \in V$ with $(u,v) \in E(G')$, $u$ and $v$ belong to the same connected component of $G[\textsc{Ext}(V)]$.
If $(u,v) \in E(G)$, this is obvious.
So suppose $(u,v) \notin E(G)$.
Since $(u,v) \in E(G') \backslash E(G)$ and $I$ is an independent set, there exists $z \in I$ such that $u,v \in N_G(z)$.
We have $z \in \textsc{Ext}(V)$ because $u,v \in V$.
Thus, $(u,z)$ and $(z,v)$ are edges of $G[\textsc{Ext}(V)]$, which implies that $u$ and $v$ belong to the same connected component of $G[\textsc{Ext}(V)]$.
\end{claimproof}

\medskip

\noindent
\textbf{Claim 2.} \textit{If $U,V \subseteq V(G)$ do not touch in $G'$, then $\textsc{Ext}(U)$ and $\textsc{Ext}(V)$ do not touch in $G$.}
\begin{claimproof}
Assume $\textsc{Ext}(U)$ and $\textsc{Ext}(V)$ touch in $G$, and we shall show $U$ and $V$ touch in $G'$.
By the assumption, there exist $u \in \textsc{Ext}(U)$ and $v \in \textsc{Ext}(v)$ such that $u \in N_G[v]$ and $v \in N_G[u]$.
If $u \in U$ and $v \in V$, then $U$ and $V$ touch in $G$ (and thus in $G'$).
Otherwise, at least one of $u$ and $v$ is in $I$.
Without loss of generality, suppose $u \in I$.
Since $u \in \textsc{Ext}(U)$, we have $N_G[u] \cap U \neq \emptyset$.
Observe that $N_G[u] \cap V \neq \emptyset$ as well.
Indeed, if $v \in V$, then $v \in N_G[u] \cap V$ and thus $N_G[u] \cap V \neq \emptyset$.
If $v \notin V$, then $v \in I$, which implies $v = u$, since $I$ is an independent set in $G$.
Thus, $N_G[u] \cap V = N_G[v] \cap V \neq \emptyset$.
Now $N_G[u]$ intersects both $U$ and $V$.
But $N_G[u]$ is a clique in $G'$, which implies $U$ and $V$ touch in $G'$.
\end{claimproof}

Now we are ready to prove $G' \in \mathcal{G}_H$.
Recall that $H = K_h^\bullet$.
Write $X \subseteq V(H)$ as the set of the $h$ clique vertices of $H$ and $Y \subseteq V(H)$  as the set of the $\binom{h}{2}$ subdivision vertices of $H$.
Assume $G'$ contains $H$ as an induced minor.
We shall construct an $H$-induced-minor of $G$, which contradicts the fact $G \in \mathcal{G}_H$.
For each vertex $x \in V(H)$, denote by $G_x'$ as the connected subgraph of $G'$ corresponding to $x$ in the $H$-induced-minor of $G'$.
Without loss of generality, we may assume that each $G_x'$ is minimal in the sense that no vertex can be removed from $G_x'$ while keeping the minor structure.
With this minimality assumption, for each $y \in Y$ with $N_H(y) = \{a,b\} \subseteq X$, $G_y'$ has to be an induced path $(v_0,v_1,\dots,v_r)$ in $G'$ such that $v_0$ (resp., $v_r$) is adjacent to $G_a'$ (resp., $G_b'$) and $v_1,\dots,v_r$ (resp., $v_0,v_1,\dots,v_{r-1}$) are not adjacent to $G_a'$ (resp., $G_b'$), in $G'$.
In particular, for every $y \in Y$ and $x \in X$, at most one vertex of $G_y'$ is adjacent to $G_x'$ in $G'$.

\medskip

\noindent
\textbf{Claim 3.} \textit{$V(G_y') \cap I = \emptyset$ for all $y \in Y$.}
\begin{claimproof}
As observed above, $G_y'$ is an induced path in $G'$.
Assume $V(G_y') \cap I \neq \emptyset$ and $v \in V(G_y') \cap I$.
Then $v$ must be an endpoint of $G_y'$.
Indeed, if $v$ is an internal vertex of $G_y'$, then the two neighbors of $v$ in $G_y'$ are contained in $N_G(v) = N_{G'}(v)$.
But $N_G(v)$ is a clique in $G'$, which contradicts the fact that $G_y'$ is an induced path in $G'$.
Thus, $v$ is an endpoint of $G_y'$.
Let $N_H(y) = \{a,b\} \subseteq X$.
Without loss of generality, assume $v$ is adjacent to $G_a'$.
If $v$ is the only vertex of $G_y'$, then $v$ is also adjacent to $G_b'$.
As $N_G(v)$ is a clique in $G'$, this would imply that $G_a'$ and $G_b'$ are adjacent in $G'$, which is impossible as $(a,b) \notin E(H)$.
Therefore, $v$ has a neighbor $v'$ in $G_y'$.
Now $V(G_a') \cap N_G(v) \neq \emptyset$ and $v' \in N_G(v)$, which implies $v'$ is adjacent to $G_a'$ in $G'$.
However, this contradicts the fact that at most one vertex of $G_y'$ is adjacent to $G_a'$.
As a result, $V(G_y') \cap I = \emptyset$.
\end{claimproof}

For each $x \in X$, define $P_x = \textsc{Ext}(V(G_x'))$.
Since $G_x'$ is connected for all $x \in X$, $G[P_x]$ is connected for all $x \in X$ by Claim~1.
Then for each $y \in Y$, define $P_y = \textsc{Ext}(V(G_y')) \backslash (\bigcup_{x \in X} P_x)$.
We show that $G[P_y]$ is connected for all $y \in Y$.
Clearly, $G'[P_y]$ is connected, since $G_y'$ is connected and all vertices in $P_y \backslash V(G_y')$ are adjacent to $V(G_y')$ in $G'$.
So it suffices to show for all $u,v \in P_y$ such that $u,v \in N_G(z)$ for some $z \in I$, $u$ and $v$ lie in the same connected component of $G[P_y]$.
Let $z \in I$ and $u,v \in P_y \cap N_G(v)$, where $u \neq v$.
Note that $z \in \textsc{Ext}(V(G_y'))$.
Thus, if $z \notin \bigcup_{x \in X} P_x$, then $z \in P_y$ and we are done.
Otherwise, $z \in P_x$ for some $x \in X$.
We observe that this is impossible.
Indeed, if $z \in P_x$, then $N_G[z] \cap V(G_x') \neq \emptyset$.
Since $N_G[z]$ is a clique in $G'$ and $u,v \in N_G[z]$, we see that both $u$ and $v$ are adjacent to $G_x'$ in $G'$, which contradicts our previous observation that at most one vertex in $G_y'$ is adjacent to $G_x'$.
As a result, $G[P_y]$ is connected for all $y \in Y$.

In the $H$-induced-minor of $G$ we are going to construct, each vertex $x \in V(H)$ just corresponds to the (connected) subgraph $G[P_x]$.
To see this truly gives us an $H$-induced-minor, we need to show for any different vertices $x,y \in V(H)$,
\begin{enumerate}[(i)]
    \item $P_x \cap P_y = \emptyset$,
    \item $P_x$ and $P_y$ are adjacent in $G$ iff $(x,y) \in E(H)$.
\end{enumerate}
If $(x,y) \notin E(H)$, then $V(G_x')$ and $V(G_y')$ do not touch in $G'$.
Claim~1 then directly implies $P_x$ and $P_y$ do not touch in $G$, because $P_x = \textsc{Ext}(V(G_x'))$ and $P_y = \textsc{Ext}(V(G_y'))$.
As such, conditions (i) and (ii) hold in this case.
So assume $(x,y) \in E(H)$.
Then one of $x,y$ belongs to $X$ and the other belongs to $y$.
Without loss of generality, suppose $x \in X$ and $y \in Y$.
Condition (i) directly follows from the construction of $P_y$.
To see condition (ii), note that $V(G_x')$ and $V(G_y')$ are adjacent in $G'$.
Thus, there exist $u \in V(G_x')$ and $v \in V(G_y')$ satisfying $(u,v) \in E(G')$.
By Claim~3, $v \notin I$.
Since $v \in V(G_y') \backslash I$, $v \notin \textsc{Ext}(V(G_a'))$ for any $a \in X$, which implies $v \in P_y$.
Also, $u \in V(G_x') \subseteq P_x$.
Now if $(u,v) \in E(G)$, then $P_x$ and $P_y$ are adjacent in $G$.
Otherwise, $(u,v) \in E(G') \backslash E(G)$ and thus $u,v \in N_G(z)$ for some $z \in I$.
We then have $z \in \textsc{Ext}(V(G_x')) = P_x$.
This implies $P_x$ and $P_y$ are adjacent in $G$, because $(z,v) \in E(G)$.

Based on the above discussion, we see that $G$ contains $H$ as an induced minor, where each vertex $x \in V(H)$ corresponds to the (connected) subgraph $G[P_x]$.
This contradicts the assumption $G \in \mathcal{G}_H$ and completes the proof.
\end{proof}

%% file: sparsity.tex

The second structural lemma we need is the following, which essentially states that when applying the \textsc{Torso} operation to a bounded-degeneracy core $M$ of an induced-minor-free graph, the increase of the number of edges is linear in $|M|$.

\begin{lemma} \label{lem:sparsity}
Let $H$ be a fixed graph.
For any graph $G \in \mathcal{G}_H$ and any $d$-core $M \subseteq V(G)$ of $G$ such that $G[M]$ is $\Delta$-degenerate, we have
\begin{equation*}
    |E(\textsc{Torso}(G,M))| - |E(G)| = O((d+\Delta) \cdot |M|).
\end{equation*}
The constant hidden in $O(\cdot)$ only depends on $H$.
\end{lemma}
\begin{proof}
Write $E^* = E(\textsc{Torso}(G,M)) \backslash E(G)$.
Our goal is to show $|E^*| = O((d+\Delta) \cdot |M|)$.
To this end, our main idea is to construct a random subset $E_0 \subseteq E^*$ such that $\mathbb{E}[|E_0|] = O(|M|/(d+\Delta))$ and $\Pr[e \in E_0] = \Omega(\frac{1}{(d+\Delta)^2})$ for each edge $e \in E^*$, which will imply $|E^*| = O((d+\Delta) \cdot |M|)$.

Since $G[M]$ is $\Delta$-degenerate, we have an orientation of the edges of $G[M]$ for which the out-degree of every vertex is at most $\Delta$ in $G[M]$.
We denote by $\varGamma(v)$ the set of out-neighbors of a vertex $v \in M$ in $G[M]$.
Then $|\varGamma(v)| \leq \Delta$ for all $v \in M$.

Let $R \subseteq M$ be a random subset obtained by sampling each vertex $v \in M$ independently with probability $p = \frac{1}{d+\Delta}$.
Define $I= \{v \in R: \varGamma(v) \cap R = \emptyset\}$.
Note that $I$ is an independent set in $G[M]$ (and thus in $G$), because for any $u,v \in I$, we have $u \notin \varGamma(v)$ and $v \notin \varGamma(u)$.
Furthermore, for each $v \in M$, $\Pr[v \in I] \leq p(1-p)^\Delta = \Theta(p)$, since $|\varGamma(v)| \leq \Delta$.
By the linearity of expectation, it follows that $\mathbb{E}[|I|] = \Theta(p|M|)$.
Write $G' = G - (M \backslash I)$.
Define $G_I'$ as the graph with $V(G_I') = I$ in which two vertices $u,v \in I$ are connected by an edge if there exists a connected component $C$ of $G'-I$ such that $N_G(V(C)) \cap I = \{u,v\}$.
Also, define $E_0 = E(G_I')$.
Since $G' \in \mathcal{G}_H$, Lemma~\ref{lem:minorfree} implies that $G_I'$ is $H$-minor-free and thus $|E_0| = O(|I|)$ by the standard extremal bound for excluded minors~\cite{Thomason2001}.
Therefore, $\mathbb{E}[|E_0|] = O(\mathbb{E}[|I|]) = O(p|M|)$.

Consider an edge $e = (u,v) \in E^*$.
Since $e \in E(\textsc{Torso}(G,M))$ and $e \notin E(G)$, there exists a connected component $C$ of $G-M$ such that $u,v \in N_G(V(C))$.
Since $M$ is a $d$-core of $G$, we have $|N_G(V(C))| \leq d$.
Note that $e \in E_0$ if both of the following events happen:
\begin{enumerate}[(i)]
    \item $u,v \in R$,
    \item $z \notin R$ for all $z \in \varGamma(u) \cup \varGamma(v) \cup (N_G(V(C)) \backslash \{u,v\})$.
\end{enumerate}
Since $|\varGamma(u) \cup \varGamma(v)| \leq 2 \Delta$ and $|N_G(V(C))| \leq d$, the probability that both events happen is at least $p^2 (1-p)^{(d+2\Delta)} = \Omega(p^2)$, which implies $\Pr[e \in E_0] = \Omega(p^2)$.
Thus, $\sum_{e \in E^*} \Pr[e \in E_0] = \Omega(p^2 |E^*|)$.
On the other hand, we have $\sum_{e \in E^*} \Pr[e \in E_0] = \mathbb{E}[|E_0|] = O(p|M|)$.
As a result, it follows that $|E^*| = O(|M|/p) = O((d+\Delta) \cdot |M|)$.
\end{proof}

As a consequence of the above two lemmas, we are able to prove the following structural result for induced-minor-free graphs.

\begin{corollary} \label{cor:structure}
Let $H$ be a fixed graph and $G \in \mathcal{G}_H$.
If there exists a $d$-core $M$ of $G$ such that $G[M]$ is $\Delta$-degenerate and $\mathbf{tw}(G - M) \leq t$, then $\mathbf{tw}(G) = O(t + \sqrt{(d+\Delta) \cdot |M|})$.
\end{corollary}
\begin{proof}
Write $G^+ = \textsc{Torso}(G,M)$ for convenience.
Thanks to Lemma~\ref{lem:torsotw}, it suffices to prove that $\mathbf{tw}(G^+[M]) = O(\sqrt{(d+\Delta) \cdot |M|})$.
We first show that $|E(G^+[M])| = O((d+\Delta) \cdot |M|)$.
Since $G[M]$ is $\Delta$-degenerate, $|E(G[M])| = O(\Delta |M|)$.
Furthermore, we have
\begin{equation*}
    |E(G^+[M])| - |E(G[M])| \leq |E(G^+)| - |E(G)| = O((d+\Delta) \cdot |M|),
\end{equation*}
which follows from Lemma~\ref{lem:sparsity}.
Hence, $|E(G^+[M])| = O((d+\Delta) \cdot |M|)$.

Next, we show that $G^+[M] \in \mathcal{G}_{H'}$ for some graph $H'$ depending on $H$.
Let $G'$ be the graph obtained from $G$ by contracting each connected component of $G-M$ to a single vertex.
We have $G' \in \mathcal{G}_H$.
Note that $I = V(G') \backslash M$ is an independent set in $G'$.
Therefore, Lemma~\ref{lem:indeptorso} implies that $\textsc{Torso}(G',M) \in \mathcal{G}_{H'}$ for some $H'$ depending on $H$.
Note that $\textsc{Torso}(G',M) - I = G^+[M]$.
As such, $G^+[M] \in \mathcal{G}_{H'}$.
Lemma~\ref{lem:separator} then implies $\mathbf{tw}(G^+[M]) = O(\sqrt{|E(G^+[M])|}) = O(\sqrt{(d+\Delta) \cdot |M|})$.
As a result, $\mathbf{tw}(G) = O(t + \sqrt{(d+\Delta) \cdot |M|})$ by Lemma~\ref{lem:torsotw}.
\end{proof}